\documentclass{emulateapj}
\usepackage{graphicx}
\usepackage{natbib}
\usepackage{amsmath}
\usepackage{amssymb}

\usepackage{csz} 
\usepackage{abrev} 
\usepackage{sidecap}
\usepackage{color}

\shortauthors{Christensen et al.}

\begin{document}

\title{Tracing Outflowing Metals in Simulations of Dwarf and Spiral Galaxies}

\author{Charlotte R. Christensen}
\affil{Physics Department, Grinnell College, 1116 Eighth Ave., Grinnell, IA 50112, United States}
\email{christenc@grinnell.edu}

\author{Romeel Dav\'e}
\affil{School of Physics and Astronomy, The University of Edinburgh, Royal Observatory Edinburgh, Blackford Hill, Edinburgh, EH9 3HJ, United Kingdom}

\author{Alyson Brooks}
\affil{Department of Physics and Astronomy, Rutgers University, the State University of New Jersey, 136 Frelinghuysen Road, Piscataway, NJ 08854-8019, United States}

\author{Thomas Quinn}
\affil{Astronomy Department, University of Washington, 3910 15th Ave NE, Seattle, WA 98195-0002, United States}

\author{Sijing Shen}
\affil{Institute of Theoretical Astrophysics, University of Oslo, Postboks 1029, Blindern, 0315 Oslo, Norway}

\begin{abstract}
\csznote{
We analyze the metal accumulation in dwarf and spiral galaxies by following the history of metal enrichment and outflows in a suite of twenty high-resolution simulated galaxies. These simulations agree with the observed stellar and gas-phase mass-metallicity relation, an agreement that relies on large fractions of the produced metals escaping into the CGM. For instance, in galaxies with Mvir ~ 1e9.5 -- 1e10 solar masses, we find that about ~ 85
}
We analyze the metal accumulation in dwarf and spiral galaxies by following the history of metal enrichment and outflows in a suite of twenty high-resolution simulated galaxies.
These simulations agree with the observed stellar and gas-phase mass-metallicity relation, an agreement that relies on large fractions of the produced metals escaping into the CGM.
For instance, in galaxies with $M_{vir} \sim 10^{9.5} - 10^{10} \Msun$, we find that about $\sim$ 85\% of the available metals are outside of the galactic disk at $z = 0$, although the fraction decreases to a little less than half in Milky Way-mass galaxies. 
In many cases, these metals are spread far beyond the virial radius.
We analyze the metal deficit within the ISM and stars in the context of previous work tracking the inflow and outflow of baryons.
Outflows are prevalent across the entire mass range, as is reaccretion.
We find that between 40 and 80\% of all metals removed from the galactic disk are later reaccreted.
The outflows themselves are metal enriched relative to the ISM by a factor of 0.2 dex because of the correspondence between sites of metal enrichment and outflows.
As a result, the metal mass loading factor scales as $\eta_{metals} \propto v_{circ}^{-0.91}$, a somewhat shallower scaling than the total mass loading factor.
We analyze the simulated galaxies within the context of analytic chemical evolution models by determining their net metal expulsion efficiencies, which encapsulate the rates of metal loss and reaccretion.
We discuss these results in light of the inflow and outflow properties necessary for reproducing the mass-metallicity relation.

\end{abstract}

\section{Introduction}\label{sec:intro}

Galaxies evolve through a balance between gas accretion and outflows.
Cosmological accretion of gas from in the inter-galactic media enables the continued growth of halos \citep{Nelson2013}, and metal-poor cold-gas accretion has been detected through absorption on the outskirts of galaxies \citep{Kacprzak2012, Bouche2013,Crighton2013}.
Additionally, reaccretion of previously ejected material provides continued fuel and can easily dominate over cosmological accretion in galaxies with halo masses \textgreater$10^{11} - 10^{12} \Msun$  \citep{Oppenheimer10}. 
This material exists as metal-enriched gas in the circumgalactic media (CGM) prior to its reaccretion \citep{Cheung2016}.

Meanwhile, gas loss from galaxies is accomplished through feedback-driven outflows.
Such outflows are ubiquitous in high-redshift star forming galaxies and local starburst galaxies \citep[e.g.][]{Heckman1990,Pettini2001,Shapley2003,Martin05, Weiner09,Steidel10,rubin13}, and both semi-analytic models and simulations have found them  necessary to reproduce key observations such as the stellar mass-halo mass relation \citep{Henriques2013,scannapieco11,Stinson12,Hopkins2013, White15}.
Together with accretion, outflows set the baryonic content within the disk and regulate star formation \citep{Dave11b, Lilly2013, Dekel2014, Christensen2016}.

In addition to regulating the baryonic content in galaxies, outflows are key to establishing their metal content.
For example, comparisons of the total metals within the interstellar media and stellar disk to the total mass of metals produced predict that 20-25\% of metals remain in the stars and ISM of Milky Way-mass galaxies \citep{Peeples2014} and 6\% remained within the stars and ISM of a dwarf galaxy \citep{McQuinn2015}.
As a result, outflows are a leading candidate to regulate the metallicity within the disks of galaxies and establish the mass-metallicity relation (MZR)~\citep[e.g.][]{tremonti04,Finlator08,Ma2016} and its second-parameter dependences on star formation rate and gas content~\citep{Dave11b}.  
The amplitude and slope of the MZR can be explained by the tendency of outflows to reduce the effective yield and by the greater efficiency of outflows in removing material from low-mass halos in combination with their reduced star formation efficiency.
While analytic models can explain the MZR by parameterizing metal inflow and outflow efficiencies \citep{	
erb08, Spitoni2010, Peeples2011, Dave2012, Lilly2013} these models generally do not account for the reaccretion of metal-enriched material.
Additionally, many of these models assume that outflows share the same metallicity as the ISM, while observations show evidence for metal enrichment compared to the ISM \citep{Chisholm2016}.
Understanding the rates of reaccretion and the relative enrichment of outflows are key to understanding the source of the MZR.

A corollary to the outflow-driven metal depletion of disk material is the redistribution of metals to the CGM and beyond.
Since metals originate primarily in the stellar disks of galaxies, their presence throughout the CGM provides a tracer of the history of inflows and outflows.
In particular, strong transport of metals by galactic outflows is indicated by the large, oxygen-rich halos surrounding present-day  \citep{Tumlinson2011,Prochaska2011a} and high redshift \citep{Lehner2014} star forming galaxies.
Observations of metal line absorption around dwarf galaxies \citep{Bordoloi2014}, around the Andromeda galaxy \citep{Lehner2015}, and throughout the intergalactic medium (IGM) \citep[e.g.][]{Cooksey2013, DOdorico2013, Shull2014} provide additional evidence for outflow-driven enrichment.
On the theoretical side, simulations generally require strong outflows from stellar feedback in order to reproduce the rapidly-advancing observations of metal lines around galaxies \citep[e.g][]{Stinson12, Shen13, Ford13a,Hummels13,Suresh2015}.
Metal-line absorption observations also provide a range of constraints to the thermal and dynamical state of the CGM and indicate a primarily-bound, multiphase CGM with photoionized and/or collisionally ionized gas embedded in a hotter low-density medium \citep[for a review, see][]{Tumlinson2017}.

Galaxy formation simulations can both provide the provenance of metals in the disk and halo and establish the history of metal accretion and outflow.
Simulations thus far have primarily focused on examining total baryonic mass loss and reaccretion \citep[e.g.][]{Oppenheimer10,Woods2014,Christensen2016,Muratov2015,AnglesAlcazar2016}.
They have tended to converge on mass loading factors with mass scalings between those expected for momentum and energy-conserving winds.
Simulations also have tended to agree that recycling of material is common, fuels late-time star formation \citep{Oppenheimer10,Woods2014}, and modifies the angular momentum profile \citep{Brook11b,Ubler2014, Christensen2016}.
However, the fate of outflowing gas, including the rates and timescales of outflow reaccretion, are highly model dependent, illustrating the importance of largely unexamined processes happening within CGM.
The examination of metals within and surrounding galaxy halos can help delineate between models by, for instance, tracing the eventual distribution of stellar-enriched material.
As an example of this type of theoretical investigation, \citet{Shen2011} found satellite progenitors and nearby dwarf galaxies to be the source of 40\% of metals within 3 $R_{vir}$ of a $z = 3$ progenitor of a Milky Way-mass halo.
In a different investigation, \citet{Muratov2017} found high recycling rates of metals at early times and in low-mass galaxies, leading to similar metallicities of inflowing and outflowing material within the central halos.
In contrast, outflows from their L$^*$ galaxies at low redshift were very weak, leading the to the accumulation of metals within stars.
We expand upon these types of studies by following the accumulation of metals within galaxies by tracing gas flows. 

Following on the work of \citet{Christensen2016},  we use a suite of galaxy-formation simulations to quantify the cycle of metal production, loss, and accretion over two and a half-orders of magnitude in virial mass.
By tracking the history of smoothed particle hydrodynamic gas particles, we identify instances of accretion and ejection, determine the eventual location of the metals produced by the galaxy, and measure the metallicity of the outflows.
In \S\ref{sec:methods}, we present the suite of simulations and describe the analysis.
Results are presented for the redshift zero metal census (\S\ref{sec:census}) and metal distribution (\S\ref{sec:metaldist}), the history of metal cycling (\S\ref{sec:metalhist}), the metallicity of outflows  (\S\ref{sec:outflowmetals}), and the metal mass loading factor and (\S\ref{sec:massloading}).
These results are discussed in light of the MZR and other work (\S\ref{sec:discuss}), and our conclusions are presented in \S\ref{sec:conclude}.

\section{Simulation and Analysis}\label{sec:methods}

\begin{deluxetable*}{lccccccc}   
\tablecaption{
Properties of the set of galaxies at z = 0
}
\tablecolumns{9} 
\tablewidth{0pt} 
\tablehead{
\colhead{Sim.} &
\colhead{Softening} &
\colhead{Gas Particle} &
\colhead{Halo ID} &
\colhead{Virial Mass} &
\colhead{Gas Mass} &
\colhead{Stellar } &
\colhead{$V_f$}\\
\colhead{Name}  &
\colhead{Length} &
\colhead{Mass}	 &
\colhead{		}   &					
\colhead{		} &
\colhead{in $R_{vir}$ } & 
\colhead{Mass} &
\colhead{		} \\ 
\colhead{ } &
\colhead{[pc]} &	
\colhead{[$\Msun$]} &	
\colhead{ } &
\colhead{[$\Msun$]}	 &
\colhead{[$\Msun$]}	 &
\colhead{[$\Msun$]} 	  &
\colhead{[km/s]}}
\startdata
				& (1)			& (2)				& (3)				& (4)					& (5)						& (6)					& (7) 		\\\hline \hline
h799 			& 87			& $3.3 \times 10^3$	& 1$^{2,3,5}$		& $2.4\times 10^{10}$	& $1.4\times 10^{9}$			& $1.4\times 10^{8}$		& 55			\\ 
				&			&				& 4				& $6.8\times 10^{9}$		& $4.1\times 10^{7}$			& $1.8\times 10^{7}$		& 33			\\ 
				&			&				& 6				& $4.4\times 10^{9}$		& $3.9\times 10^{7}$			& $3.5\times 10^{6}$		& 27			\\ 
\hline
h516	 			& 87			& $3.3 \times 10^3$	& 1$^{1,2,3,5}$		& $3.8\times 10^{10}	$	& $2.3\times 10^{9}$			& $2.5\times 10^{8}$		& 67			\\ 
			 	&			&				& 2		 	   	& $1.5\times 10^{10}	$	& $3.7\times 10^{8}$			& $8.1\times 10^{7}$		& 34			\\ 
\hline
h986			 	& 170		&  $2.7 \times 10^4$	& 1$^{3,5}$  		& $1.9\times 10^{11}$	& $1.7\times 10^{10}$		& $4.5\times 10^{9}$		& 103		\\ 
			 	&			&				& 2		 	   	& $5.9\times 10^{10}	$	& $3.2\times 10^{9}$  		& $1.2\times 10^{9}$	  	& 77			\\ 
				&			&				& 3		 	   	& $3.8\times 10^{10}	$	& $2.4\times 10^{9}$ 		& $4.6\times 10^{8}$	  	& 76			\\ 
				&			&				& 8		 	   	& $1.1\times 10^{10}	$	& $6.4\times 10^{7}$		 	& $4.0\times 10^{7}$	  	& 35			\\
				&			&				& 15				& $4.4\times 10^{9}	$	& $8.7\times 10^{7}$ 	 	& $6.2\times 10^{6}$		& 29			\\
				&			&				& 16				& $3.2\times 10^{9}	$	& $3.0\times 10^{7}$ 		& $2.3\times 10^{6}$		& 27			\\
\hline
h603			 	& 170		& $2.7 \times 10^4$	& $1^{3,5}$ 		& $3.4\times 10^{11}$	& $3.1\times 10^{10}	$		& $7.8\times 10^{9}$		& 115		\\ 
			 	&			&				& $2^3$		 	& $1.0\times 10^{11}	$	& $6.1\times 10^{9}$		  	& $3.8\times 10^{9}$	  	& 75			\\
				& 			&				& 3		 	   	& $2.9\times 10^{10}	$	& $1.8\times 10^{8}$		 	& $3.9\times 10^{8}$  	& 50			\\
\hline
h258			 	& 170		& $2.7 \times 10^4$	& $1^{3,4}$		& $7.7\times 10^{11}	$	& $5.6\times 10^{10}$	  	& $4.5\times 10^{10}	$  	& 182		\\
			 	&			&				& 4		 	   	& $1.1\times 10^{10}	$	& $1.4\times 10^{8}$  	  	& $5.9\times 10^{7}	$  	& 43			\\
\hline
h285			 	& 170		& $2.7 \times 10^4$	& $1^{3}$			& $8.8\times 10^{11}	$	& $6.3\times 10^{10}$ 	  	& $4.6\times 10^{10}	$  	& 164		\\
			 	&			&				& 4		 	   	& $3.4\times 10^{10}	$	& $1.2\times 10^{9}$		  	& $3.9\times 10^{8}$  	& 64			\\
				&			&				& 9		 	   	& $1.2\times 10^{10}	$	& $3.1\times 10^{8}$		  	& $5.4\times 10^{7}$  	& 52			\\
\hline
h239			 	& 170		& $2.7 \times 10^4$	& $1^{3}$		 	& $9.1\times 10^{11}$	& $8.1\times 10^{10}$    		&$4.5\times 10^{10}$  	& 165		

\enddata
\tablenotetext{1}{Appears in \citet{Christensen12}.}
\tablenotetext{2}{Appears in \citet{Governato12}.}
\tablenotetext{3}{Appears in \citet{Munshi12}.}
\tablenotetext{4}{Appears in \citet{Zolotov12}.}
\tablenotetext{5}{Appears in \citet{Christensen12a}.}

\end{deluxetable*}

We use cosmological simulations of seven individual volumes to follow the history of metals in  twenty field
galaxies with final virial masses between $10^{9.5}$ to $10^{12}\Msun$.
These are the same set of simulations analyzed in \citet{Christensen2016}.
An overview of them is given below and a description with greater detail may be found in \citet{Brooks2017}.

These simulations were computed using the $N$-Body + SPH  code, {\sc gasoline} \citep{Wadsley04}.
{\sc gasoline} is an SPH extension to the parallel, gravity-tree based $N$-Body Code {\sc pkdgrav}
\citep{stadel01}.
The simulations assume a $\Lambda$CDM cosmology using WMAP3 \citep{Spergel07} parameters: $\Omega_0$=0.24, $\Lambda$=0.76, h=0.73, $\sigma_8$=0.77. 
In order to achieve high-resolution while including the cosmological context, we used the ``zoom-in'' volume renormalization technique \citep{katz93}.
The final sample includes galaxies selected from a medium resolution 50$^3$ Mpc$^3$ volume and a higher resolution 25$^3$ Mpc$^3$ volume.
In the high (medium) resolution simulations, the force spline softening lengths are $\epsilon=87$ (170)~pc and the particle masses for the dark matter, gas, and stars (at their formation)
are, respectively, 1.6 (13)$\times 10^4$, 3.3 (27.0)$\times$10$^3$, and 1.0 (8.0)$\times$10$^3$M$_{\odot}$.

{\sc gasoline} follows non-equilibrium abundances of H (including H$_2$) and He species.
Photo-ionization and heating rates are based on a redshift-dependent cosmic ultraviolet background (Haardt \& Madau 2005)\footnote{ Haardt \& Madau (2005) refers to an unpublished updated version of \citet{Haardt96}, specified in {\sc cloudy} \citep{Ferland98} as ``table HM05.''},
while $\Hmol$ dissociation is based on the Lyman-Werner radiation produced by nearby star particles \citep{Christensen12}.
H and He cooling channels include collisional ionization \citep{Abel97}, $\Hmol$ collisions, radiative recombination \citep{Black81,
VernerANDFerland96}, photoionization, bremsstrahlung, and line cooling \citep{Cen92}
Metal line cooling rates are calculated from {\sc cloudy} \citep[version 07.02;][]{Ferland98} models based on the gas temperature, density, metallicity, and cosmic UV background under the assumptions of ionization equilibrium and optically thin gas.

Star formation proceeds stochastically according to \begin{equation}
 p = \frac{m_{gas}}{m_{star}}(1 - e^{-c^* \frac{X_\Hmol}{X_{\Hmol} + X_{\mathrm{HI}}} \Delta t /t_{dyn}})
\end{equation}
where $p$ is  the probability of a gas particle spawning a star particle in a time step $\Delta t$, $m_{gas}$ is the mass of the gas particle, $m_{star}$ is
the mass of the potential star, $c^*=0.1$ is the star forming efficiency,
$X_\Hmol$ and $X_\mathrm{HI}$ are the mass
fractions of the particle in the form of $\Hmol$ and HI, respectively, and $t_{dyn}$ is the dynamical time.
The dependency on the $\Hmol$ abundance ensures that star formation happens in dense ($\rho \gtrsim 10$ amu cm$^{-3}$), cold gas; however, star formation is technically allowed in any gas particle with densities greater than 0.1 amu cm$^{-3}$ and temperatures less than 10$^3$ K.

Energy from SN II is distributed to the surrounding gas particles according to the ```blastwave'' approach \citep{Stinson06}, assuming a \citet{Kroupa93} initial mass function (IMF) and the canonical 10$^{51}$ ergs per supernova.
In this sub-grid recipe, the cooling of feedback-affected particles is disabled for the theoretical lifetime of a hot, low-density
shell produced during the momentum-conserving phase of the supernova remnant \citep{McKee77}.
This recipe differs from many other recipes \citep[e.g.][]{SpringelANDHernquist03a,Dave11,scannapieco11} in that no momentum kick is added to the particles; the particle remains hydrodynamically coupled to the rest of the simulation, and the feedback depends only on the local gas properties.
We do not include a separate model for other forms of stellar feedback, such as radiation pressure, \citep[e.g.][]{Hopkins2013,Stinson13} that help drive a galactic wind through additional momentum transfer or by making the gas more responsive to the supernova feedback.
Instead, this blastwave recipe represents the total stellar feedback from young stars.

We follow the production and distribution of oxygen and iron separately.
Metals are returned to the ISM both through type I and II SNe and stellar winds.
For SN II, metals are distributed to the same gas particles as the feedback energy, assuming the production rates from \citet{Raiteri96} with yields from \citet{Woosley1995}.
SN Ia are calculated to occur using the rates from \citet{Raiteri96}.
Each SN Ia produces 0.63 $\Msun$ iron and 0.13 $\Msun$ oxygen \citep{Thielemann1986}, which are transferred to the nearest gas particles.
Energy from SN Ia is also distributed to the gas particles within the smoothing kernel; however, cooling is not disabled for these particles as it is for SN II.
Mass is returned to the ISM by stellar winds using mass loss rates from \citet{Weidemann87}.
This mass is distributed to gas particles within the smoothing sphere of the star particle, assuming the same metallicity as the star particle.

Metals are further distributed throughout the gas through diffusion \citep{Shen10}.
In this model, based on \citet{Smagorinsky1963}, subgrid turbulent mixing is treated as a shear-dependent diffusion term.
As a result, the highest diffusion rates are calculated for shearing flows. 
Instances of compressive or purely rotating flows result in no diffusion.
A tunable parameter, called the metal diffusion coefficient, is used to scale the strength of diffusion.
For these simulations, a conservative value of 0.01 was chosen for the metal diffusion coefficient.

\subsection{Post-processing Analysis}
Individual halos are selected from the snapshots of the simulations during post-processing.
We used {\sc amiga's halo finder} \citep{Knollmann2009,Gill2004},\footnote[1]{{\sc amiga's halo finder} is available for download at http://popia.ft.uam.es/AHF/Download.html} in which areas of over-density are identified using grid hierarchy and then gravitationally unbound particles are iteratively removed from the prospective halos.
The virial radius, $R_{vir}$, is defined such that the average halo density is a multiple of the background density.
This multiple evolves with redshift but is approximately equal to 100 times the critical density at $z = 0$.
The main progenitor of each galaxy was traced back in time from redshift zero through the creation of a merger tree, and was defined at each snapshot to be the halo that contained the majority of particles from the subsequent snapshot.

\subsubsection{Inflow/Outflow Identification}
\label{sec:parttrace}
After the main progenitor halo had been identified at each snapshot, we identified all instances of accretion and outflow.
To do this, gas particles are classified in each snapshot as being part of the disk, within the halo, and outside the galaxy.
Accretion events and ejection are found by identifying the instances when particle move from one classification to another.
The time resolution of this tracking comes from the spacing between snapshots, which were $\sim 100$ Myrs for all the simulations.
As a result, we are unable to measure accretion and ejection at a time resolution less that 100 Myr and it is possible for accretion and ejection events to be missed if a particle leaves and reenters the galaxy between timesteps.

We use the following criteria for classifying particles into different components.
Gas particles are defined as being in the halo if they are within the virial radius.
They are defined as within the disk if they 1) have density $> 0.1$ amu cm$^{-3}$, 2) have temperature $< 1.2 \times 10^4$ K, and 3) are less than 3 kpc from the plane of the disk.
These disk parameters were chosen to thermodynamically select for the ISM and to spatially eliminate cool gas in accreting satellites.

Particles are classified as being accreted onto the halo each snapshot they go from not being part of a halo to being included in it. 
Similarly, particles are classified as being accreted onto the disk each step they are included in disk after not having been included in it during the previous snapshot.
Particles may, therefore, be accreted onto the disk or halo multiple times.
As may be expected, particles are identified as leaving the disk each time they move from disk to halo.
However, we also further divide the ejected material between material that is merely heated and material that passes more stringent constraints for ejecta.
This division is necessary as it is common for gas particles to be heated by SN to temperatures greater than $1.2 \times 10^4$ K but then to quickly radiate their energy away without having a substantial effect on the disk dynamics.
Therefore, we define the following classifications, with each later classification a subset of the previous ones.
\begin{itemize}
\item Material {\em removed} from the disk includes all gas that is classified as disk material in one step and outside of the disk in the next step.
\item Material {\em ejected} from the disk includes those particles removed from the disk that become dynamically unbound from the baryonic disk. Specifically, a particle is defined to be ``unbound from the disk" at any snapshot after leaving a disk if its velocity exceeds the escape velocity for a mass equal to the sum of the ISM and stellar mass located at the center of the galaxy. This classification is most similar to what is generally referred to as outflows.
\item Material {\em expelled from the halo} includes all particles that are classified as in the disk in one instance and at any later point travel beyond the virial radius. Note that we are focusing our analysis on gas that was ever part of the ISM and so do not include in our analysis gas that was only ever part of the halo prior to leaving it.
\end{itemize}
Together, these classifications allow us to determine all gas particles that were ever part of the halo or disk and their history of accretion and outflow.

One additional complication in tracking the flow of metals in and out of galaxies is that metal diffusion is included in the simulation. 
As a result, metals may pass from within the disk to the halo through diffusion, as opposed to an SPH particle leaving the disk.
Therefore, it is best in these circumstances to think of the SPH particles as tracer particles of the metal distribution that pick up on bulk motion while missing some of the small scale diffusion.
Based on comparing the mass of metals within disk and halo at each time step to that expected from particle tracking, we are able to account for $\gtrsim 90$\% of the metal movement from disk to halo and back.
Throughout this paper, we note those aspects of the analysis where diffusion could further effect the results and specify how we addressed it.

\section{Results}\label{sec:results}
\subsection{Mass-Metallicity Relation}\label{sec:mzr}

We assess our simulated galaxies with reference to the observed MZR.
These simulations were previously shown to be consistent with the redshift zero gas-phase MZR, stellar mass-halo mass relation, and the baryonic and $i$-band Tully Fisher relation \citep{Christensen2016}.
Galaxies generated with a previous version of {\sc gasoline} have also been shown to demonstrate an evolving gas-phase MZR \citep{brooks07}.
Here, we expand our analysis to show our simulated galaxies along the gas-phase MZR at multiple redshifts and the redshift zero stellar MZR.

\begin{figure*}
\begin{center}$
\begin{array}{cc}
\includegraphics[width=0.5\textwidth]{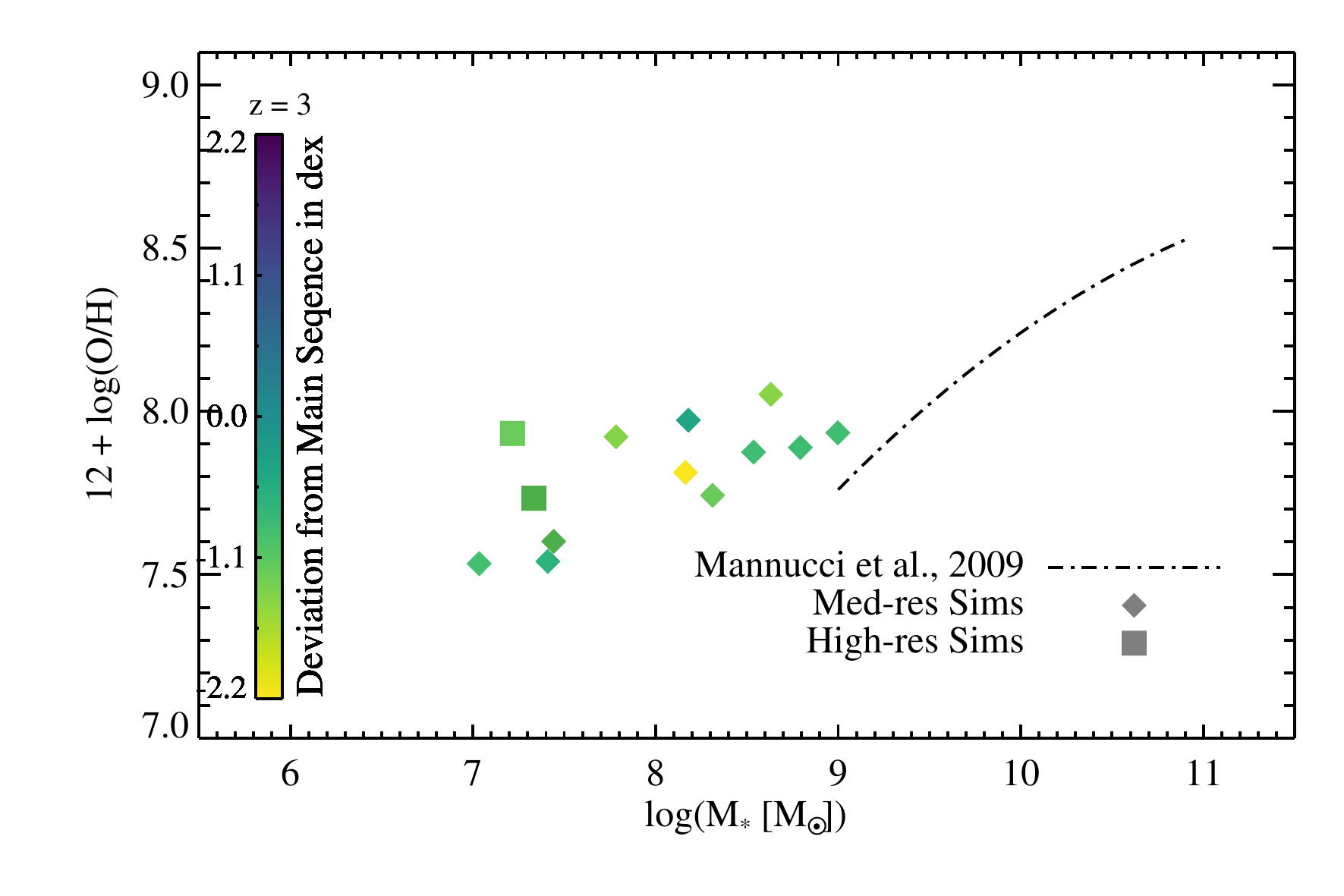}
\includegraphics[width=0.5\textwidth]{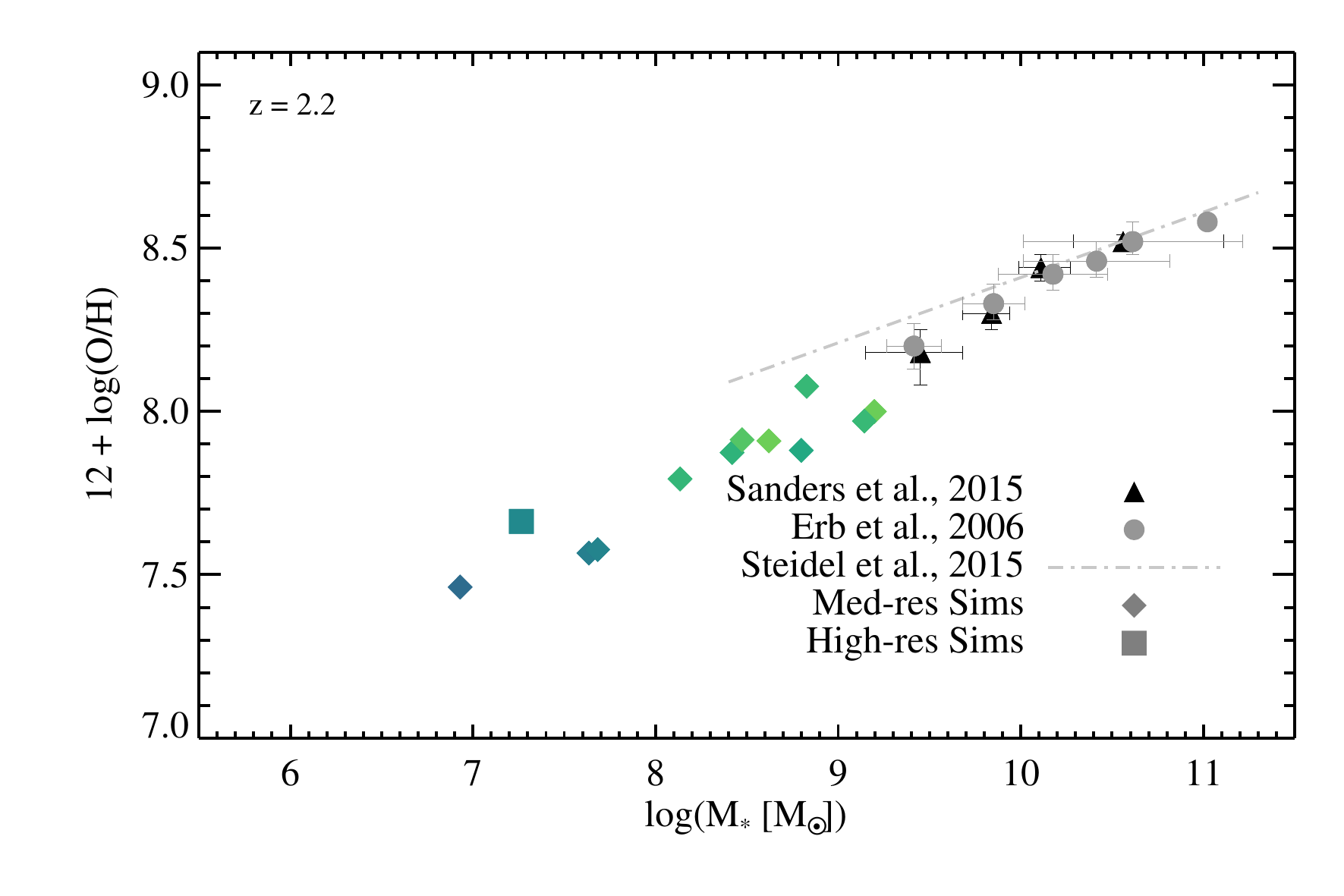}\\
\includegraphics[width=0.5\textwidth]{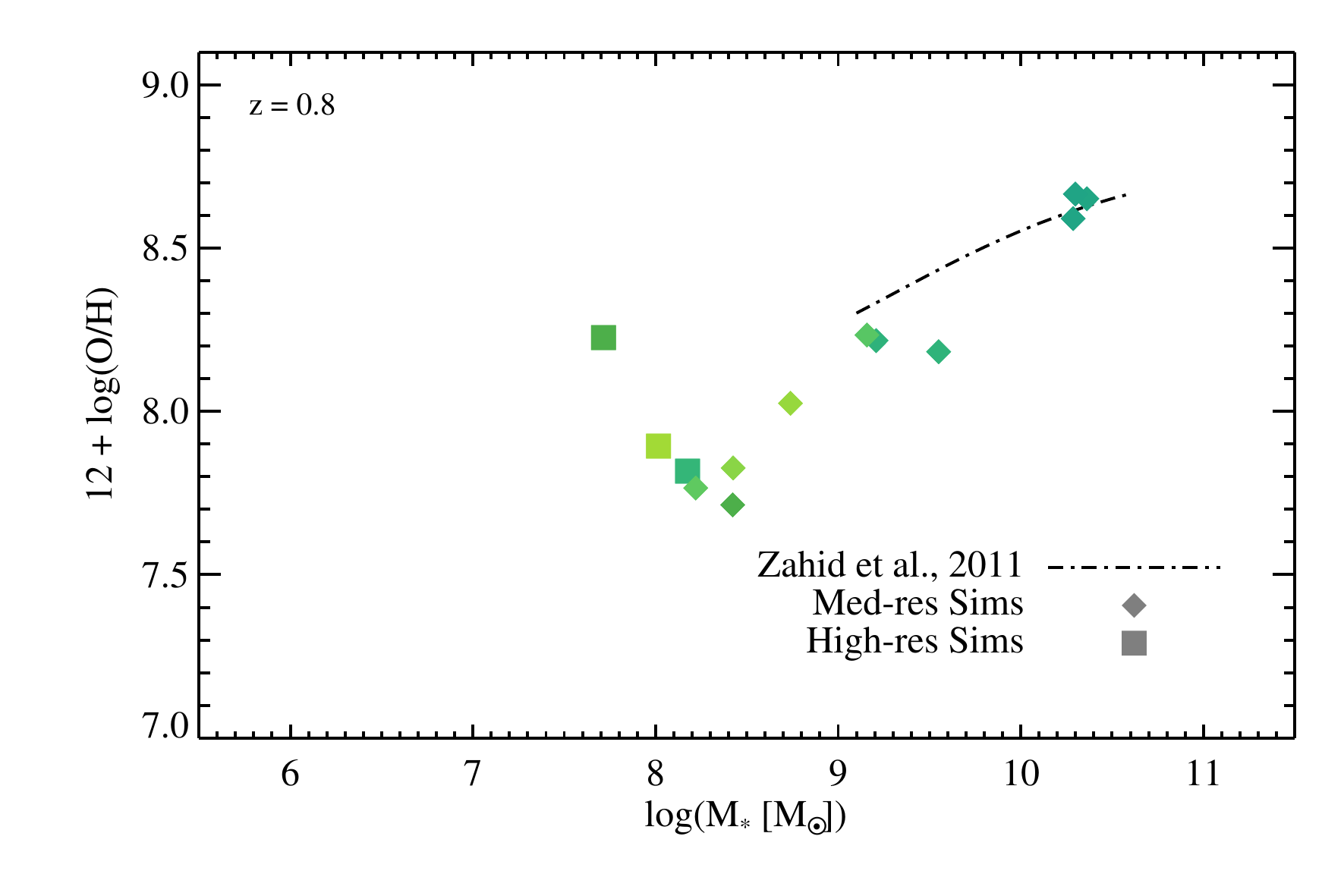}
\includegraphics[width=0.5\textwidth]{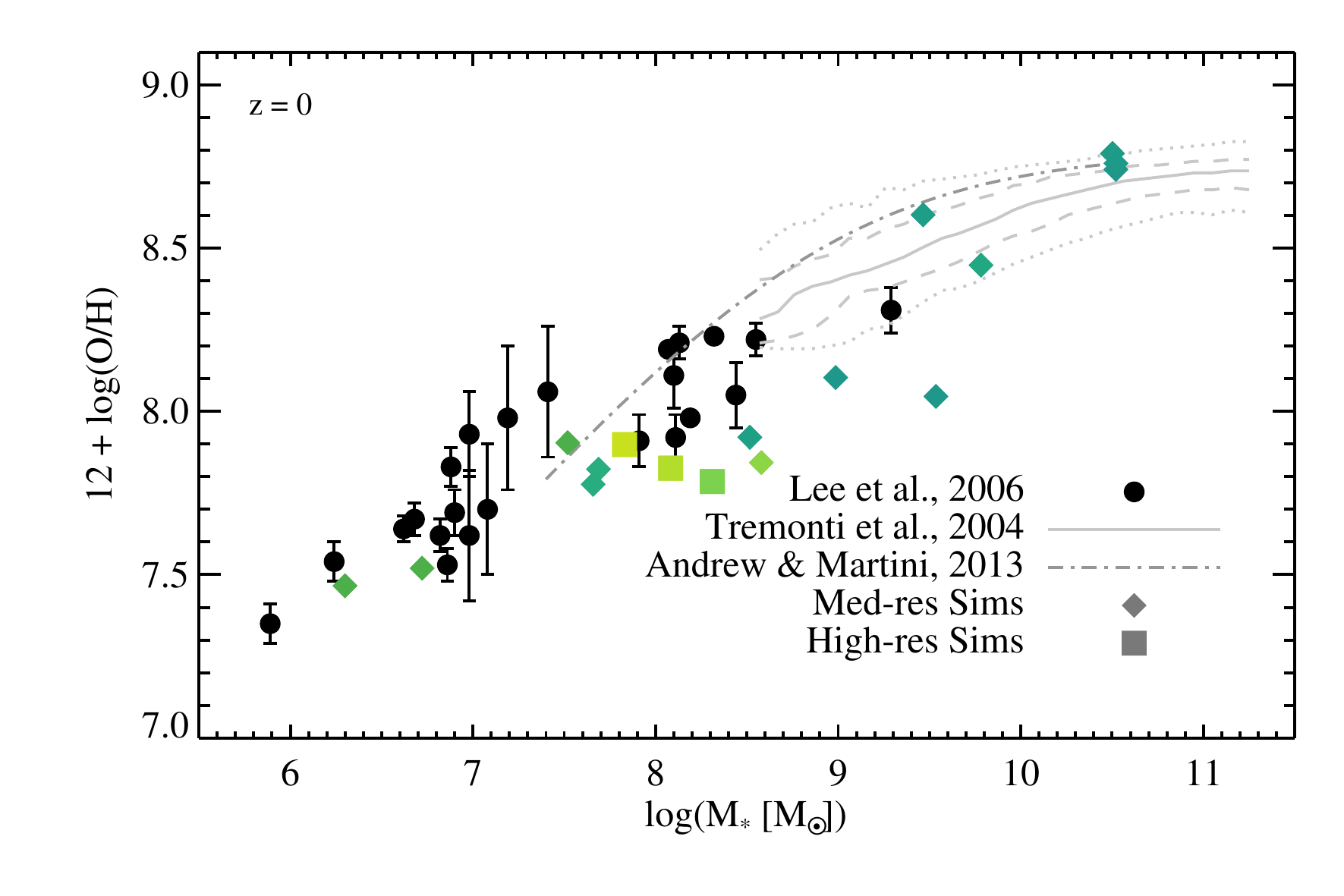}
\end{array}$
\end{center}
\caption[Evolution of the MZR]
{ 
The evolving MZR for simulated galaxies (colored symbols) at redshifts 3.0, 2.3, 0.8 and 0 compared to observations (gray-scale lines and symbols). 
Stellar masses of simulated galaxies were determined from mock-photometric observations.
The higher resolution simulations are represented by squares and the lower resolution by diamonds.
The top-left panel compares the redshift 3.0 simulations to the observed sample of Lyman Break galaxies from \citet{Mannucci2009} (black dot-dashed line).
The top-right panel compares the redshift 2.3 simulations to data from \citet{Erb06} (gray circles), \citet{Steidel2014} (gray dot-dashed line) and \citet{Sanders2015} (black triangles) based on N2 calibrations \citet{Pettini2004}.
The bottom left panel compares the redshift 0.8 simulations to data from \citet{Zahid2011} (black dot-dashed line), adjusted to N2 calibration following \citet{Kewley08}.
The bottom right panel compares the redshift zero simulations to data from \citet{tremonti04}, adjusted to N2 calibration (gray lines, solid showing the median values, dashed enclosing 68\%, and dotted enclosing 95\% of the galaxies). 
Additional redshift zero observations are shown from \citet{AndrewsMartini2013} (gray dot-dashed line) and \citet{Lee06} (black circles).
Colors represent the amount by which galaxies deviate from the galaxy main sequence, measured in dex.
Fits to the observed main sequence for each of the redshift ranges are taken from \citet{Salim2007} ($z = 0$), \citet{Whitaker2014} ($z$ = 0.8 and 2.2), and \citet{Santini2017} ($z$ = 3).
}
\label{fig:mzr}
\end{figure*}

Figure~\ref{fig:mzr} compares the gas-phase oxygen abundances at z = 3, 2.3, 0.8, and 0 to observed values.
In order to best mimic observations, stellar masses are determined from broad band magnitudes using {\sc kcorrect} \citep{Blanton07}.
Metallicities were calculated using the star formation rate-weighted average gas particle oxygen abundances.
Weighting by the star formation rate (i.e. probability of the gas particle forming a star) was chosen to mimic the measurement of metallicities in star forming regions of observed galaxies. 
The simulated galaxies show a slight increase in the normalization of the MZR with decreasing redshift.

We compare our simulated galaxies to observational data from $z \sim 3$ \citep{Mannucci2009}, $z \sim 2.3$ \citep{Erb06, Steidel2014, Sanders2015}, $z \sim 0.8$  \citep{Zahid2011}, and $z \sim 0$ \citep{tremonti04,AndrewsMartini2013,Lee06}.
These observations were made based on different diagnostics using different calibrations, and the systematic uncertainty between different metallicity diagnostics could be as much as 0.7 dex \citep{Kewley08}.
In order to demonstrate the evolution of the MZR, we show or convert to N2 calibration  \citep{Pettini2004} where possible; in particular, we use the formula in \citet{Kewley08} to transform \citet{tremonti04} and \citet{Zahid2011} to the N2 calibration.
Compared to observations, the simulations at z = 3 may have slightly too high of metallicity.
However, given the lack of overlap in the mass range and the large systematic uncertainties in observed metallicities, the simulations appear largely consistent with observations.

We looked for evidence of a second parameter dependence of the metallicity on star formation rate at a given stellar mass in our sample, by coloring the galaxies in Figure~\ref{fig:mzr} by their deviation from the star formation rate expected for their stellar mass.
Specifically, the difference between the log measured specific star formation rate and log expected star formation rate is encoded into color on a scale spanning $\pm$2 dex.
Star formation rates were calculated from the simulations by averaging over the previous million years, and were compared to the star formation rates expected from a fit to the specific star formation rate-stellar mass main sequence at the relevant redshift: \citet{Salim2007} at $z = 0$, \citet{Whitaker2014} at $z$ = 0.8 and 2.2, and \citet{Santini2017} at $z$ = 3.
It is important to recognize, though, that data for galaxies with stellar masses below $10^9 \Msun$ is rare beyond the local universe so the fits to observational data had to be extrapolated to the relevant stellar mass regime.
Our simulations do not show clear evidence for a second parameter dependence at any redshift.  
This is likely the result of our small sample size, but also may owe to the fact that most of the galaxies are in the mass regime where the star formation is bursty~\citep[as discussed in][]{FaucherGiguere2018}, and such burstiness will hide this second parameter dependence on short timescales \citep{Torrey2018}.

\begin{figure}
\begin{center}
\includegraphics[width=0.5\textwidth]{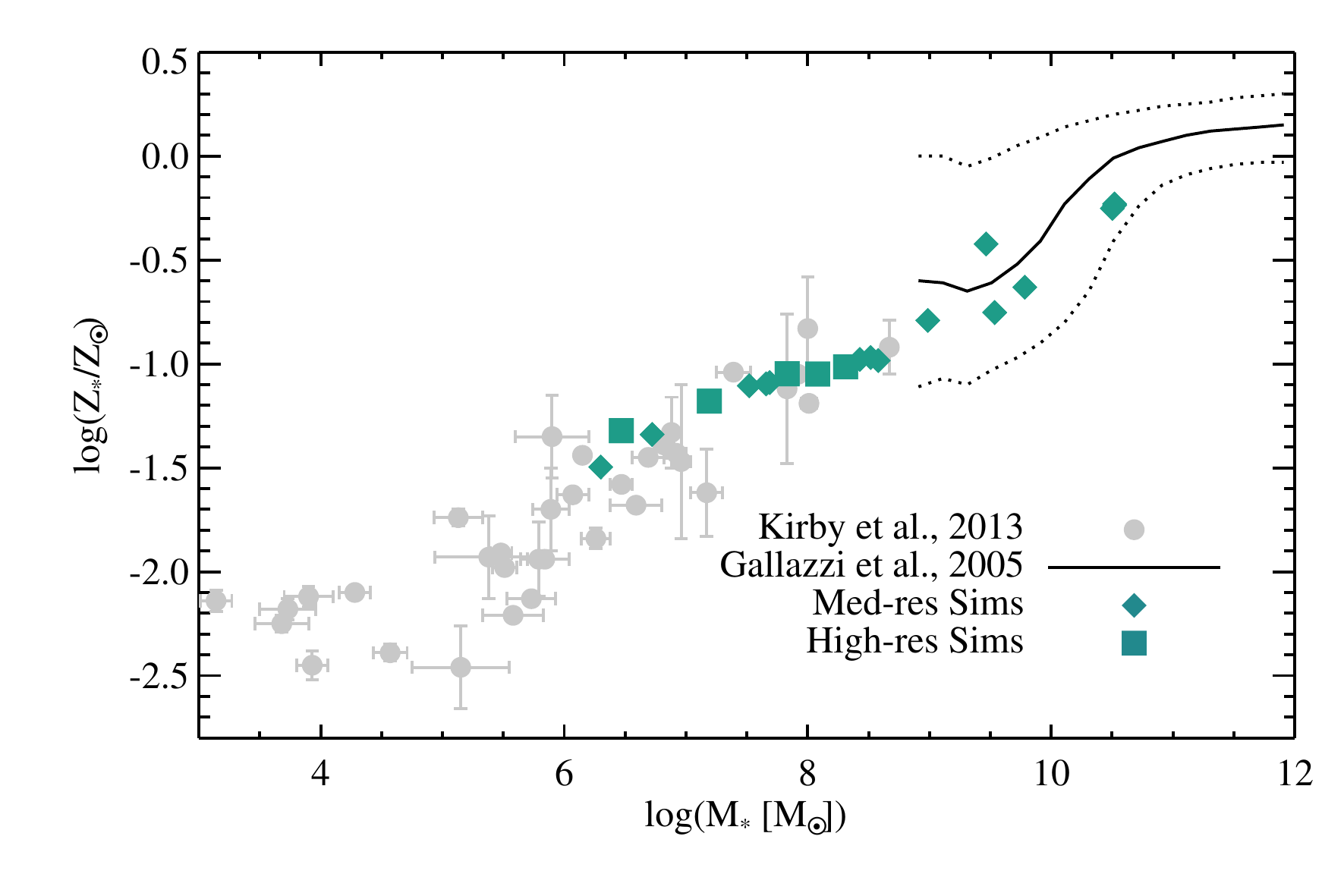} 
\end{center}
\caption[stellar mass-metallicity relation]
{
The stellar MZR for simulated galaxies (colored symbols) compared to observed values (gray-scale symbols).
Redshift zero stellar metallicities from simulations were calculated using the $K$-band weighted average metallicity of all star particles in the galaxy.
Stellar masses of simulated galaxies were determined from mock-photometric observations.
The higher resolution simulations are represented by teal filled squares and the lower resolution by diamonds.
Gray filled circles represent observational data for dwarf irregular and dwarf spheroidal galaxies from \citet{Kirby2013}.
Black lines show observational data from \citet{Gallazzi2005}; the solid line shows the median values and the dotted lines show the 16th and 84th percentile data.
The simulated galaxies appear consistent with the observed data and there is no distinction between the two different resolutions.
}
\label{fig:stellarmzr}
\end{figure}

Figure \ref{fig:stellarmzr} shows the simulated and observed z = 0 MZR for stellar metallicities.
As for the gas-phase MZR, stellar masses are calculated using {\sc kcorrect}.
In order to better compare with observations, stellar metallicities are weighted by the $K$-band luminosity of the star particle.
However, this weighting introduces almost no change to the average metallicity.
The simulations are compared to observational data from \citet{Gallazzi2005} and \citet{Kirby2013}.
We find that the simulations are consistent with the observations over 4.5 orders of magnitude, indicating that an appropriate mass of metals are retained within the stars, in addition to the ISM.
We further compare the fraction of metals present in different phases of the galaxy in the following section.

\subsection{Metal Census}\label{sec:census}
Observed galaxies shows a substantial deficit in the metals contained within their disks (including both stars and the atomic and molecular component of their ISM). 
For example, the isolated dwarf galaxy Leo P was found to contain only 5\% of all the metals produced by its stars in its disk (1\% in the form of stars) \citep{McQuinn2015}.
On the more massive side, \citet{Peeples2014} found that in galaxies with stellar masses between 10$^{9.2} \Msun$ and 10$^{11.6} \Msun$ approximately 20 -- 25\% (with an uncertainty range between 10 and 40\%) of the metals produced by the stars remained in the disks of the galaxies.
This metal deficit is a companion to the ``Missing Baryon Problem;" like the missing baryons, these metals are presumed to largely be contained within the CGM.
Here we examine the metal census for our population of galaxies.
Since we have excluded satellites (as defined by Amiga Halo Finder) from our sample, our analysis focuses on metals lost for reasons other than stripping by a more massive halo.

\begin{figure}
\begin{center}
\includegraphics[width=0.5\textwidth]{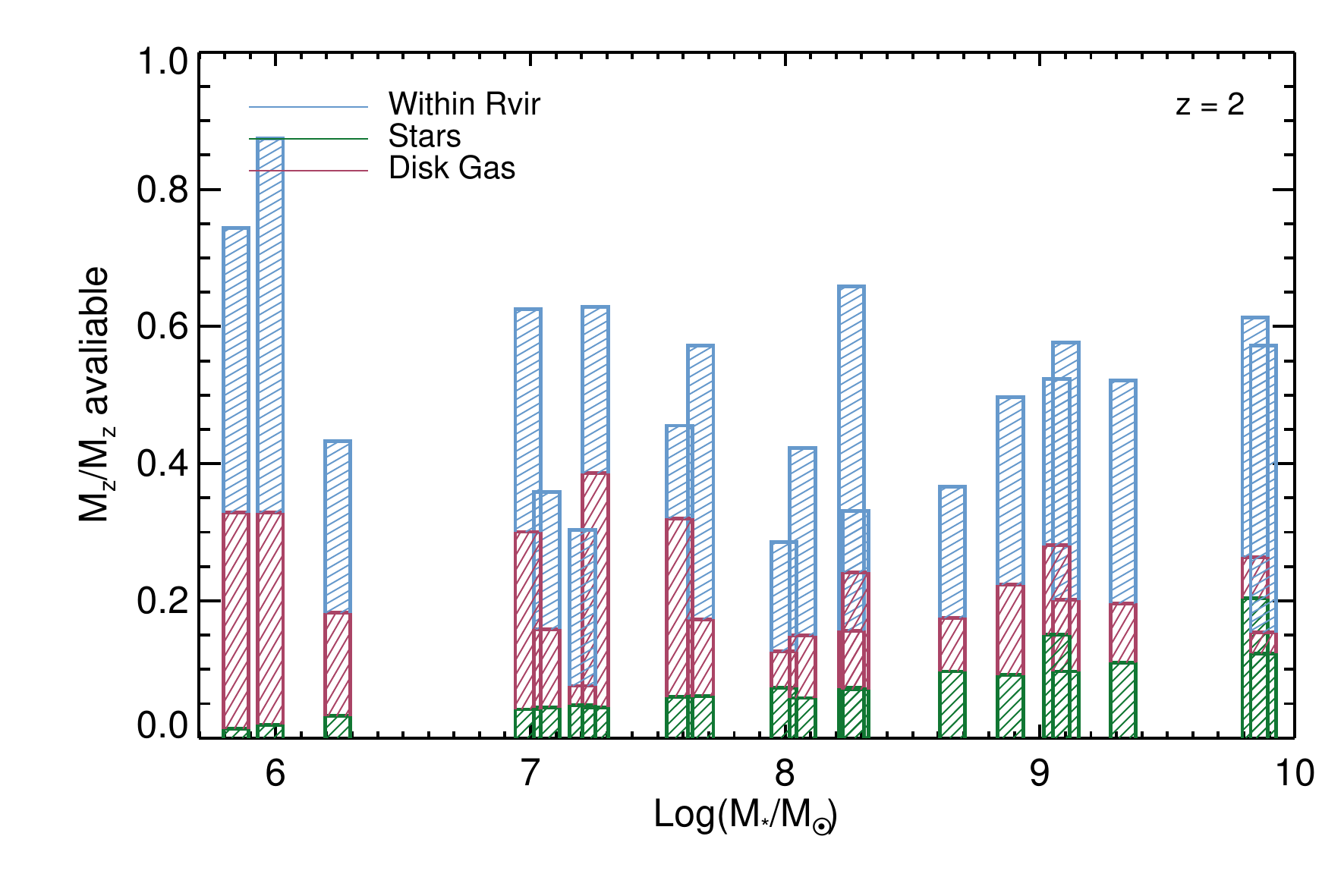}
\includegraphics[width=0.5\textwidth]{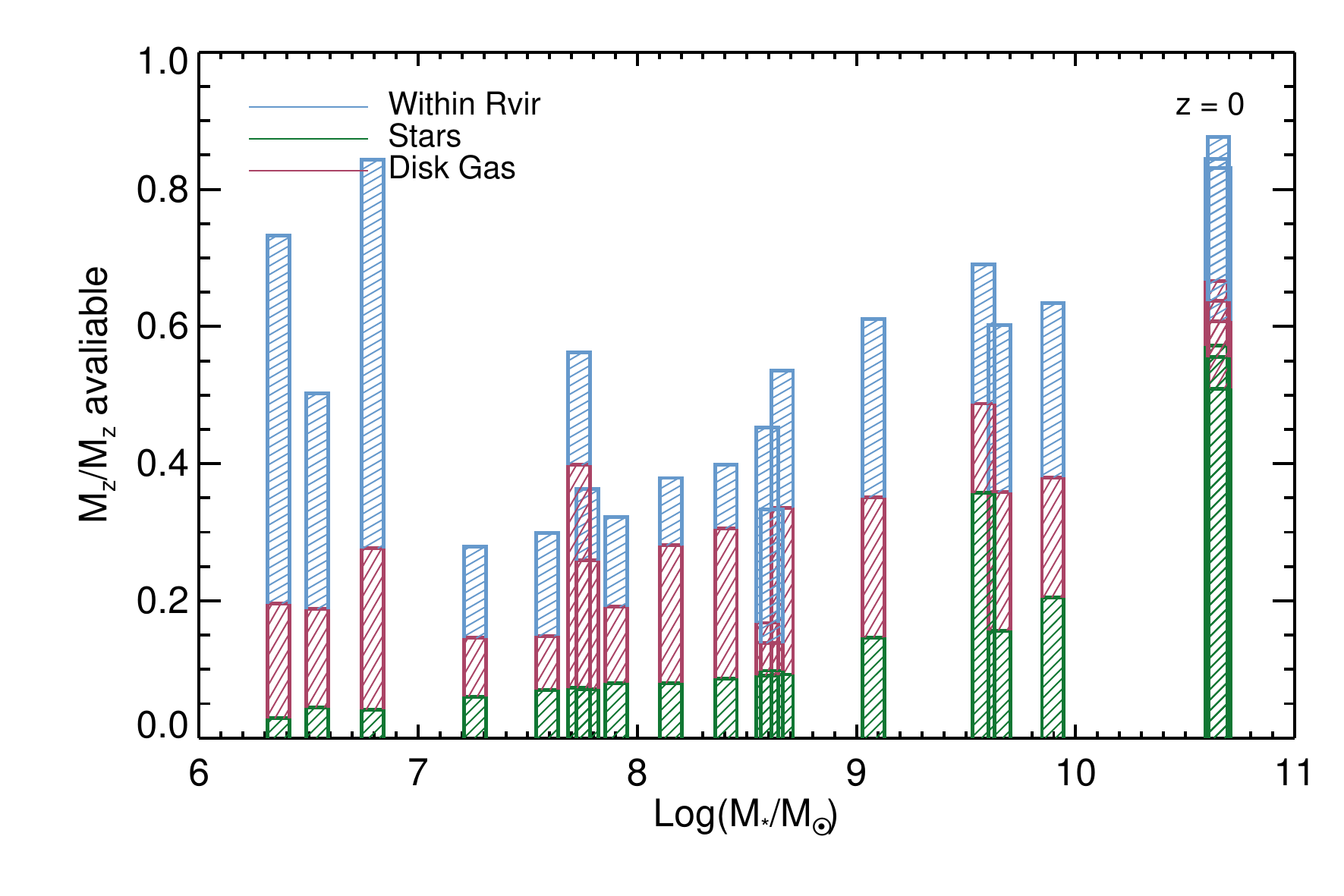} 
\end{center}
\caption[Metal mass dispersal]
{ 
The redshift $z = 2$ and $z = 0$ location of all the metals produced by each of the galaxies. 
The metal mass contained in each component is normalized by ``M$_{\mathrm{z, available}}$," which is defined to be the sum of all metals produced by the stars within the final halo.
Each bar represents a unique galaxy with the fraction contained within stars shown in green, the fraction contained within the ISM shown in maroon and the fraction within the virial radius but not within the ISM and stars (i.e., the CGM) shown in blue. 
In part because of their larger stellar mass, more massive galaxies retain a greater fraction of their metals in stars.
The metal fraction retained within the CGM does not show a clear mass trend and, by $z = 0$, neither does the fraction retained within the ISM.
}
\label{fig:metaldist}
\end{figure}

Figure~\ref{fig:metaldist} shows the fraction of metals available to each galaxy contained within the halo ($r < R_{vir}$), the stars, and the ISM at redshifts two and zero. 
The mass of metals available is defined to be the sum of all metals produced by the stars within the final halo.
To do this, we used the same metal production models included within {\sc gasoline} for SNII and SNIa to calculate the mass of oxygen and iron produced by all star particles given their age and metallicity and assuming a \citet{Kroupa93} IMF.
Since observations do not tally the metals contained within stellar remnants, we also do not include them when calculating the mass of metals contained within stars.
Specifically, we reduced the mass of metals in each star particle by the mass fraction of that particle in the form of stellar remnants based on its age and metallicity.

We find that a substantial fraction of metals are lost from galaxies of all masses; by $z = 0$ between 35\% and 85\% of the metals were removed from the galactic disk and between 15\% and 75\% from the entire halo, as defined by the virial radius.
The metal fractions contained in stars shows a strong mass dependency with higher mass galaxies retaining a greater fraction of metals in stars.
By $z = 0$, the fraction of metals contained within the entire halo also shows some evidence for mass scaling for those halos with $M_* > 10^7 \Msun$.
Within this mass range, lower mass galaxies appear more able to remove metals through outflows, likely because of their lower gravitational potential.
This result mirrors a similar one in \citet{Christensen2016}, where 20\% of baryons ever accreted to the galaxy were retained within it at $z = 0$ for dwarf galaxies, while closer to 80\% were retained within Milky Way-mass galaxies.
However, the three galaxies with $z = 0$ stellar masses $< 10^7 \Msun$ complicate this mass trend by retaining relatively large fractions of their metals within $R_{vir}$.
These galaxies may illustrate a transition to a mass range where the low gravitational potentials that could aid metal loss are counterbalanced by incredibly low rates of star formation.
Notably, similar fractions of metals are retained within the ISM for galaxies of all masses.
Similarly, the fraction of metals retained within the CGM does not show a clear mass trend.

The metal deficit is well established even by $z = 2$, with generally $\sim$60\% or less of metals retained within the virial radius.
This result is consistent with observations indicating that the missing metals problem is already in place at z$\sim$2 \citep[e.g.][]{Pagel1999}.
In the evolution from $z = 2$ to $z = 0$ the fraction of metals retained within the lowest mass galaxies ($M_{*} < 10^{7.5}$) is reduced as outflows continue to expel metals.
In contrast, the metal mass fraction within the highest mass galaxies increases, primarily as those metals become locked into stars.
Across the entire range of galaxies, the fraction of metals within the halo gas tends to decrease over time. 

A similar analysis of numerical simulations in \citet{Muratov2017} found comparable $z = 0$ mass trends in metal mass loss.
As in ours, they found that greater fractions of the available metals were locked within stars for Milky Way-mass galaxies than for dwarf galaxies, while the fraction retained within the ISM showed a negligible mass trend.
However, higher amounts of metals were retained in the CGM, stars, and ISM in their simulations than in ours.
The difference was greatest in the Milky Way-mass galaxies  ($M_* \sim 4 \times 10^{10} \Msun$).
While we find that 50-60\% of available metals are retained in stars and 80 -- 90\% within a virial radius at a redshift of zero, they found closer to 80\% in stars and \textgreater 90\% within a virial radius.
Differences between these results were most likely due to differences in implementing feedback, as will be discussed further in \S
\ref{sec:discuss}. 

Observational constraints for halos in this mass range are limited.
Nevertheless, we draw some comparisons at both the high and the low end.
The largest survey of the fraction of metals retained in stars within dwarf galaxies are for eight dwarf spheroidal Milky Way satellites with stellar masses between $5.6 \times 10^5$ and $1.8 \times 10^7 \Msun$ \citet{Kirby2011}.
They determined that $<$1\% to 4\% of the metals produced by the stars in their sample of galaxies were retained within the stellar component (with the greatest fraction retained in the most massive dwarf), which they found suggestive of an energy driven scaling for the mass loading factor.
In comparison, we found that our four galaxies in this mass range retained a similar fraction  of their metals in stars (specifically, 3, 4, 4 and 6\% for the galaxies with masses between $2.3 \times 10^6$ and $1.8 \times 10^7 \Msun$). 
Likewise, these simulations exhibit mass loading factors with energy driven scalings \citep{Christensen2016}.
Despite this apparent agreement, it is dangerous to draw strong conclusions from this comparison, as the observed sample of dwarf spheroidal galaxies has a substantially different environment and evolutionary history than our field dwarf irregular galaxies.
In particular, tidal and/or ram pressure stripping acting on the observed satellites may have impacted the fraction of metals retained within stars by contributing to metal loss prior to the cessation of star formation.
Furthermore, we cannot compare the fraction of metals retained within the ISM for our sample to data from \citet{Kirby2011}, as dwarf spheroidal galaxies are necessarily lacking an ISM because of their satellite environment.

Observational measurements of the metal census for field dwarf galaxies are difficult to achieve because H II regions are required in addition to stellar spectroscopy.
The only currently available observational metal census for a field dwarf galaxy is for Leo P \citep{McQuinn2015}.
Leo P shows no evidence of interaction and has a metallicity consistent with a low-luminosity extension of the MZR, implying that it is a representative galaxy.
It has a stellar mass half that of our lowest mass galaxy and has a similar, although slightly smaller, metal fraction retained within its stars as our two lowest-mass galaxies \citep[3 and 4\% for the simulated dwarf galaxies and $\sim$1\% in Leo P][]{McQuinn2015}.
When comparing the fraction of metals retained within the ISM, it is important to use similar definitions in selecting ISM material.
In figure~\ref{fig:metaldist}, we use the same definition of ``disk" material as in our particle tracking code: density $> 0.1$ amu cm$^{-3}$, 2) temperature $< 1.2 \times 10^4$ K, and 3) less than 3 kpc from the plane of the disk (\S~\ref{sec:parttrace}).
However, observations generally (and in the case of Leo P, specifically), measure ISM mass through HI and, when detectable, H$_{2}$.
In low-mass halos, in particular, the difference between these two ISM definitions can be significant.
Therefore, we also calculate the fraction of metals retained within the ISM as determined by scaling gas particles within 3 kpc of the disk plane by their H I and H$_2$ content.
We calculate that our two lowest mass galaxies retained 9 and 10\% of their metals within the ISM defined this way, compared to the $\sim4$\% determined for Leo P.
This factor of two difference may imply that the simulations retain too many of their metals in their ISM.
Or it is possible that the discrepancy can be explained by the differences in stellar masses between Leo P and the simulations, and stochasticity in metal retention among dwarf galaxies.
Larger samples of both observed and simulated dwarf galaxies will be needed to draw firmer conclusions as to the consistency of the results.

For our most massive galaxies, the metals retained can be compared with observations from \citet{Peeples2014}.
The average 64\% of metals we found retained in Milky Way-mass galaxy disks is substantially greater than the $\sim$25\% measured by \citet{Peeples2014}.
This is a qualitatively similar but smaller level of disagreement than \citet{Muratov2017} had with \citet{Peeples2014}.
It is possible that this discrepancy could argue for the need of an additional form of feedback, such as AGN, in the most massive of our galaxies.
However, comparisons to the gas and stellar MZR confirm that the metallicities in our simulations are consistent with observed values (\S\ref{sec:mzr}).
Given that the metal content within the ISM and stars agrees with observations, the discrepancy with \citet{Peeples2014} almost certainly originates in differences in how the available metal mass is calculated.
While {\sc gasoline} uses the yields from \citet{Woosley1995} for SNII, \citet{Peeples2014} assumes higher yields based on other models.
Additionally, \citet{Peeples2014} assume significantly higher mass loss rates ($\sim$55\%) from simple stellar populations using a \citet{Chabrier2003} IMF than {\sc gasoline} does using a \citet{Kroupa93} IMF.
As a result, \citet{Peeples2014} calculates about four times as much metal mass available for the same stellar mass as our simulations produce. 
Assuming a higher value of $M_{z,available}$ than what is actually used in the simulations profoundly reduces the presumed fraction of metals retained in both the disk and the CGM (defined to be material within 150 kpc of the galaxy in \citet{Peeples2014}).

By changing our calculation of the metals available (and, to a much lesser extent, how we select for ISM and CGM material) to be consistent with \citet{Peeples2014}, we can compare Milky Way-mass galaxies (log$(M_*/\Msun) \sim 10.6$) and slightly lower mass spiral galaxies (9.5 $<$ log$(M_*/\Msun) < 10.0$) to measurements from \citet{Peeples2014}.
Specifically, under these assumptions, the simulated Milky Way-mass galaxies are predicted retain on average 13\% of their metals in their disk and 17\% within 150 kpc, compared to the observed $\sim 25\%$ and $> 30\%$ for those components.
Similarly, the simulated lower-mass spiral galaxies would be predicted to retain on average 8\% in their disk and 15\% within 150 kpc, compared to the observed $\sim 20\%$ and $>40\%$.
So, by changing the calculation of $M_{z,available}$ to follow the method in \citet{Peeples2014}, we move from predicting about twice as many metals retained in the CGM and disks of spiral galaxies to half as many.
Therefore, we cannot yet claim agreement with \citet{Peeples2014}, but uncertainties in metal yields also limit the ability of observations to constrain the simulations.

\subsection{Redshift Zero Distribution of Metals}\label{sec:metaldist}

\begin{figure*}
\begin{center}
\includegraphics[width=\textwidth]{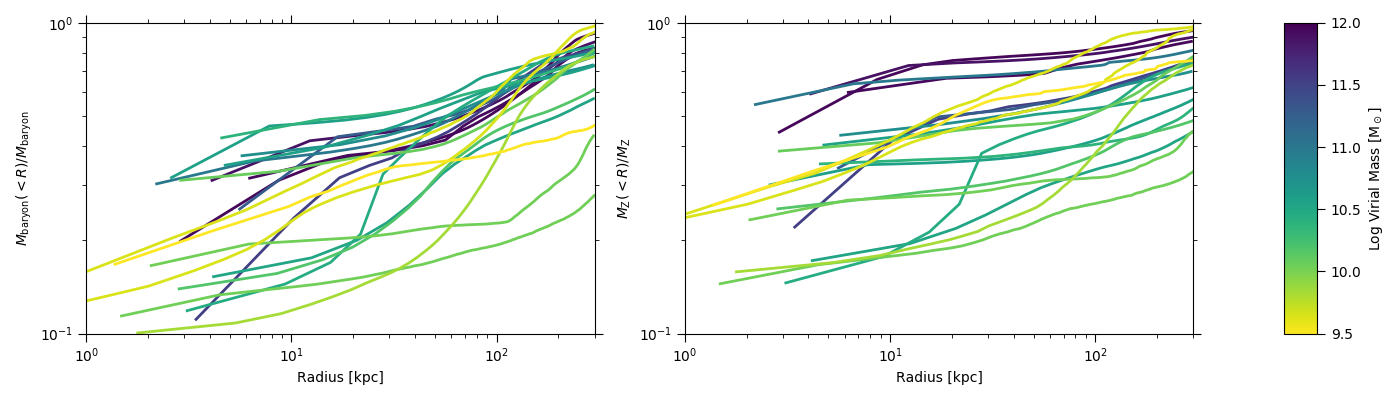} \\
\includegraphics[width=\textwidth]{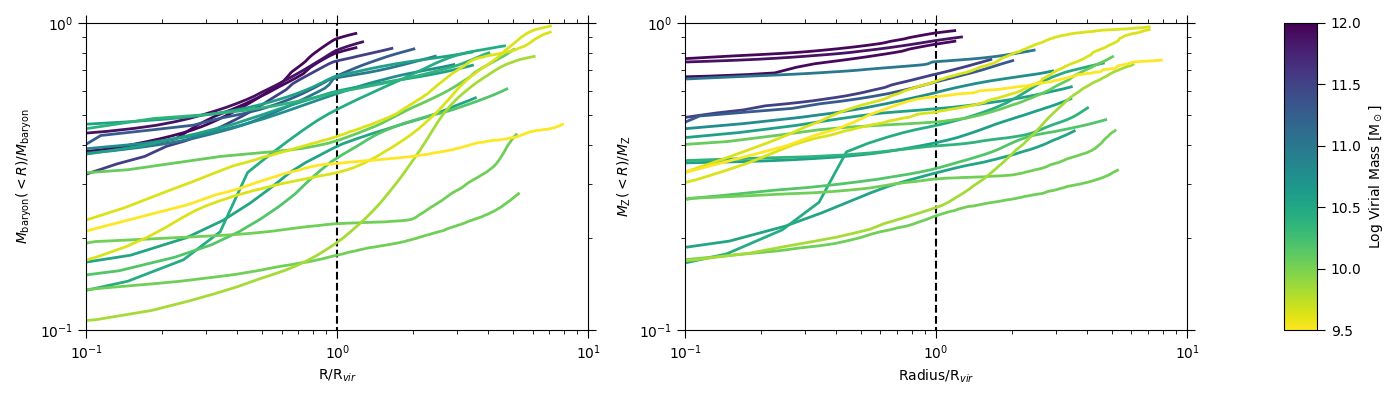} 
\end{center}
\caption[Metals as a function of radius]
{ 
Normalized, cumulative histograms of the $z = 0$ location of the mass (left) and metals (right) ever within the galaxy halo since z = 3 as a function of radius.
The top panels show the absolute distances, while in the bottom panels the distances are scaled by the virial radius of the corresponding galaxy. 
The line colors represent galaxy mass spanning from yellow (low mass) to purple (high mass).
The shape of the curves and the relationship between different galaxies is similar for the mass and metal histograms.
However, the metal mass fraction tends to be higher at small radii and flatten off more slowly at large radii.
}
\label{fig:cumhistr}
\end{figure*}

Observations of the CGM have shown metals distributed out to the virial radius \citep[e.g.][]{Tumlinson2011} and in some cases OVI has been observed out to $5 R_{vir}$ \citep{Pratt2018}, demonstrating the far reach of outflows.
Simulations have also highlight the reach of outflows, especially in low-mass galaxies.
\citet{Ma2016} found that their dwarf galaxies retained only 2 -- 20\% of their metals within a virial radius, while \citet{Shen13a} calculated that 87\% of the metals produced by a group of seven dwarf galaxies were spread over a 3$^3$ Mpc$^3$ volume, equating to a distance of $\sim$ 17.5 $R_{vir}$ of the most massive dwarf galaxy.
Here, we examine the extent of metal enrichment of the CGM by showing the metal fraction contained within a given radius.
In figure~\ref{fig:cumhistr}, the left-hand panels shows the normalized cumulative histogram of the z = 0 locations of gas or stars ever part of the galaxy halo since the start of the simulation. 
The right-hand panels weight those particles by their metal content to demonstrate the eventual distribution of metals.
The top panels show the physical distribution of the matter, while the bottom panels show the distances scaled by the virial radius.
All histograms are shown out to 300 kpc.
This distance corresponds to the largest impact parameter typically used for observations of the CGM and ensures that the analysis is within the highest resolved regions of the simulation.
Note that the normalization factor for this plot differs from the mass of metals available used in \S\ref{sec:census}, as in this plot only those metals contained within gas or star particles ever part of the main progenitor are considered.

Since metal diffusion can occur across gas particles, it is possible that the eventual location of some metals may not be the same as the gas particle they exited the halo with.
More exactly, one may consider the gas particles to be tracer particles associated with the underlying metal distribution.
Therefore, the eventual location of the gas particles follows the bulk motion of the metals at a limited resolution.
In order to avoid underestimating the mass of metals exiting the virial radius because of metal diffusion we make the following adjustment when generating figure~\ref{fig:cumhistr}.
For those particles that exit the virial radius we consider their metallicity at the time they exit, while in all other instances the redshift zero metallicity is used.

As would be expected from the substantial fractions of metals beyond the virial radius (figure \ref{fig:metaldist}), the metal enrichment continues far beyond $R_{vir}$.
The distribution of both total mass and metals relative to the virial radii shows clear trends with galaxy mass.
Metals and total mass tend to remain closer to the centers of more massive galaxies because of their larger gravitational potential.
However, this mass trend is complicated by the three lowest-mass galaxies, whose metals are less dispersed than most of the medium-mass galaxies.
The lack of mass  and metal dispersal in these smallest galaxies likely results from their extremely low star formation rates.

We find that on average about 78\% of metals are contained within the virial radii of the three most massive halos.
Galaxies with virial masses between 10$^{9.5}$ and 10$^{10.5} \Msun$ only retain on average 45\% percent of their metals within a virial radius. 
In general, the trend of increasing dispersal relative to the virial radius with decreasing virial mass is similar to that found by \citet{Ma2016}, although these simulations have slightly more metals retained within a virial radius for dwarfs and slightly fewer for Milky Way-mass galaxies than in \citet{Ma2016}.
For instance, in \citet{Ma2016} essentially all metals produced by their Milky Way-mass galaxy was retained within one virial radius (and almost all within 0.1 $R_{vir}$) while their $M_{vir} = 2.5 \times 10^{9} \Msun$ galaxy retained only 2\% of its metals within a virial radius.

\begin{figure}
\begin{center}
\includegraphics[width=0.5\textwidth]{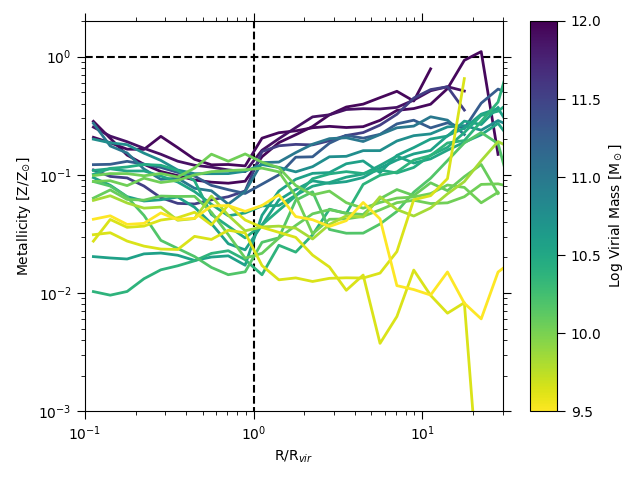}
\end{center}
\caption[Metals as a function of radius]
{ 
Metallicity of the material ever once within $R_{vir}$, calculated from the ratio of the histogram of metals to the histogram of mass.
The spatial distributions are scaled by the virial radius of each of the galaxies.
As in figure~\ref{fig:cumhistr}, the line colors represent galaxy mass spanning from yellow (low mass) to purple (high mass).
Metallicities generally decrease out to $R_{vir}$.
Only those particles ever within $R_{vir}$ are included in this analysis, so particles beyond $R_{vir}$ at $z = 0$ were likely once part of an outflow.
This selection explains the frequent rise in metallicity beyond $R_{vir}$.
}
\label{fig:cumhistratio}
\end{figure}

Metals are generally more likely to be retained close to the center of the galaxy than total mass.
This phenomenon is apparent in the shallowness of the cumulative metals histogram at very small radii.
It can be seen even more clearly by the average metallicity of the particles -- i.e., the ratio of the histogram of metals to the histogram of mass ever within the galaxy disk (figure~\ref{fig:cumhistratio}).
For almost all of the galaxies, the metallicity is relatively high at the center and drops toward the virial radius.
This relatively high metal retention can be explained by the tendency of star formation to occur throughout the galactic disk, where gas is comparatively metal-enriched, resulting in metals becoming locked into stars.
After the virial radius, the metallicities tend to rise again.
The change from inside to outside of the virial radius is a result of our selection -- since only those particles that were once within the virial radius are analyzed, those particles that are outside of the virial radius at $z = 0$ are especially likely to have been part of an outflow.
The frequent continued rise in metallicity after $R_{vir}$ results from the correlation between metal injection and supernova energy.
Those particles that travel far distances most likely received large amounts of both energy and metals.

\subsection{History of Metal Enrichment}\label{sec:metalhist}
\begin{figure*}
\begin{center}
\includegraphics[width=1\textwidth]{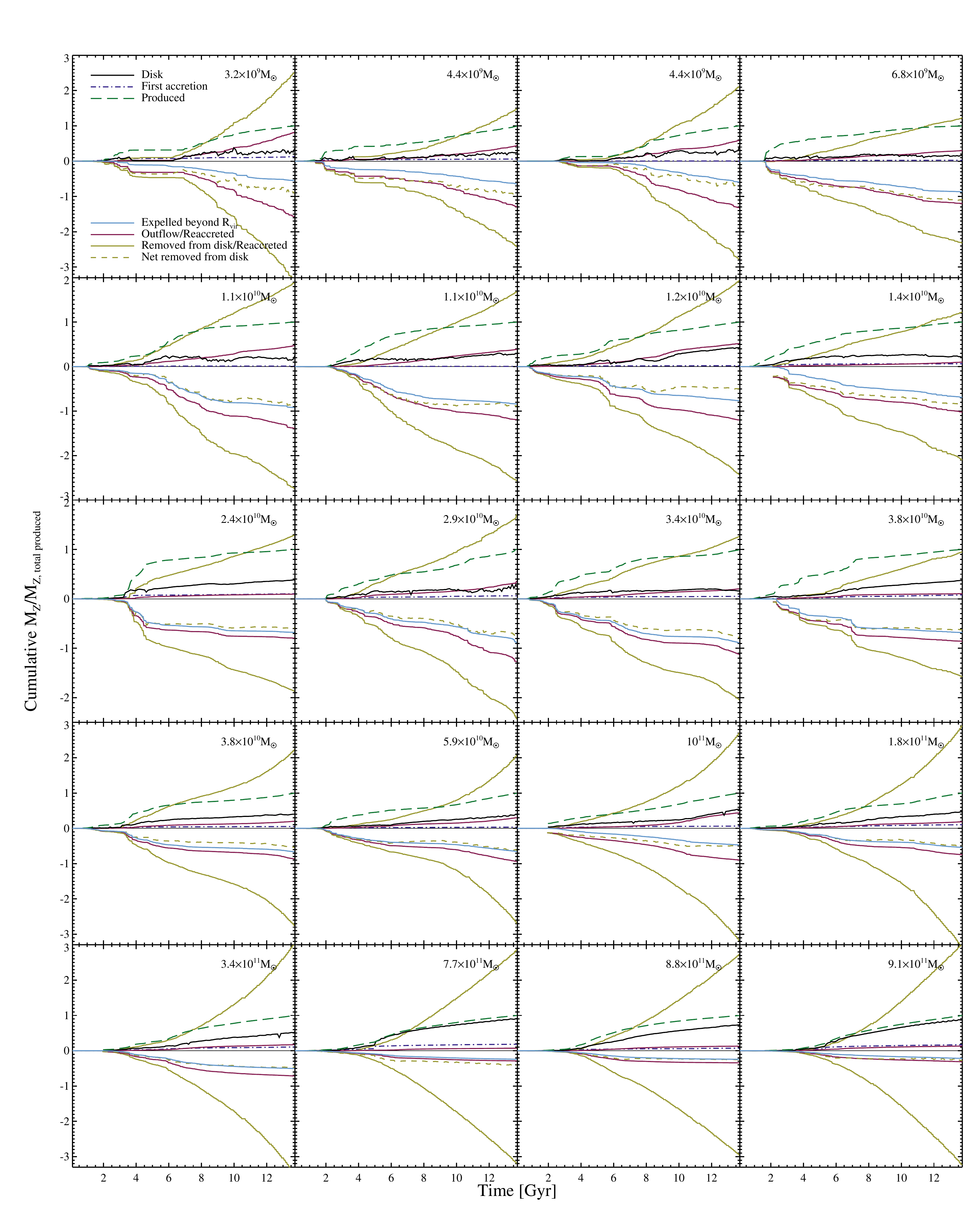} 
\end{center}
\caption[Metal enrichment history]
{ 
History of the metal build up within galaxies. All values are scaled by the total mass of metals produced by the stars in the main progenitor by $z = 0$. 
The $z = 0$ virial mass of each galaxy is listed in the upper-left corner of each panel.
The black solid line represents the metal mass contained within the ISM and stars. 
All other positive-valued lines show cumulative histograms of the mass of metals produced by stars (green long-dashed line), accreted externally (purple dot-dashed), reaccreted to the disk after being removed from it (gold), and reaccreted to the disk after being ejected (maroon).
The solid negatively-valued lines show the cumulative histograms of the mass of metals removed from the disk (gold), ejected from the disk (maroon) and expelled beyond the virial radius (light blue). 
Additionally, the net metal mass removed from the disk is shown by the dashed gold line.
The net metal mass removed was calculated by subtracting the reaccreted metal mass from the removed metal mass.
}
\label{fig:outflow_history}
\end{figure*}

The history of metal enrichment of both the CGM and ISM can be seen by chronicling pristine accretion, star formation, outflows, and reaccretion.
At any point in time, the metal content within the ISM and stars is the sum of the metals accreted from outside the halo (either in the form of gas or stars) and the metals produced by the stars within the galaxy minus the net metal loss in outflows.
This net metal loss is the total mass of metals that have left the galaxy minus the mass of metals reaccreted to it.
In figure~\ref{fig:outflow_history} we show the history of these processes and the total metal mass contained within the ISM and stars as a function of time for each of the simulated galaxies.

Metal production within the galaxy by stars is shown by the long-dashed green line.
Similar to the amount of ``metals available" generated for figure~\ref{fig:metaldist}, the metal production rates are calculated using the same enrichment models as {\sc gasoline}.
However, to find the metal production history we only consider the metals produced within the main progenitor.
To do this, the mass of metals produced between two snapshots was calculated for those star particles within the main progenitor during the latter snapshot.
Additional metals are gained through externally accreted gas and stars, frequently as part of a merger.
Rates of externally accreted metals were tallied using all gas and star particles that had previously been external to the main progenitor.
Gas particles and stars were considered accreted at the time they first entered the galactic disk; however, we determined the metal mass accreted using the metallicity of the gas particles {\em at the time they enter the halo}.
The metallicity at this earlier time was chosen to ensure that any additional metals picked up as the particle traveled through the halo were not included as external accretion.
For all halos, the metals produced by a galaxy overwhelm the metals gained through external accretion, a point that will be further quantified later in this section.

Except in the highest mass galaxies, the mass of metals contained in the ISM and stars is only a small fraction of the total metals produced and accreted from external sources, as also seen in \S\ref{sec:census}.
Therefore, it is clear that metal removal via outflows must be instrumental in setting the metal content of the galaxies.
The cumulative mass of metals removed from the disk is shown as the negatively-valued gold line.
The subset of the metals removed that achieve sufficient energy to exceed the escape velocity of the disk, what is generally considered part of an outflow and what we term as ``ejected," are shown as the negatively-valued red line.
An even smaller subset of metals is fully ``expelled" from the halo, i.e., they reach a distance farther than the virial radius.
The cumulative history of these metals is shown as the blue line.
The fraction of metals removed from the disk that satisfy either the ejected or expelled criteria depends strongly on halo mass.
The more massive the halo, the higher the energy thresholds for ejection and expulsion and the less likely a particle achieving a temperature or density sufficient to not be considered removed from the ISM will actually be part of an outflow.
An additional distinction between the material removed from the ISM and the subset that is ejected or expelled can be seen in the different shapes of the curves.
The mass of metals expelled and ejected track the mass of metals produced since both follow the star formation history (and, therefore, the history of stellar feedback) in the galaxy.
In contrast, the mass of metals lost from the disk continues to rise steeply over time as metals continually rapidly pass in and out of the disk.

Substantial amounts of metal mass are returned to the disk, indicating the importance of gas recycling.
We show the cumulative history of all metals reaccreted after leaving the disk (calculated by subtracting the externally accreted metal mass from the total accretion) as the positively valued gold line.
For illustrative purposes, we also show the cumulative history of metals reaccreted as part of particles previously ejected from the disk (red positively-valued line).
However, this quantity carries the caveat that additional metals may be reaccreted after being ejected by diffusing to other accreted particles.
Therefore, this line is a lower limit on the mass of metals actually reaccreted following an outflow.
Finally, we show the net mass of metals removed from the disk (cumulative history of metals removed minus the cumulative history of metals reaccreted after removal) as the dashed gold line.

Reaccretion is common across our entire sample.
In fact, for all galaxies there is as much metal reaccretion after removal as there is metal production; for the most massive galaxies, multiple cyclings of gas lead the metal reaccretion mass to exceed the metal production by factors of a few.
More massive galaxies tend to have slightly higher rates of reaccretion of removed metals, as shown by the difference between the {\em total} mass of metals removed and the {\em net} mass of metals removed.
Despite the prevalence of recycling, however, much of the metal mass produced remains within the CGM.
In fact, the net metal loss exceeds the total metals contained in the disk at all redshifts for any but the three most massive galaxies.
While some of these ``permanently removed" metals escape the halo, they may also remain within $R_{vir}$, as seen when the net metal loss exceeds the mass of metals expelled (in other cases where mergers result in the reaccretion of gas from outside the virial radius, the net metal loss from the disk may be less than the total mass of metals expelled). 

\begin{figure}
\begin{center}
\includegraphics[width=0.5\textwidth]{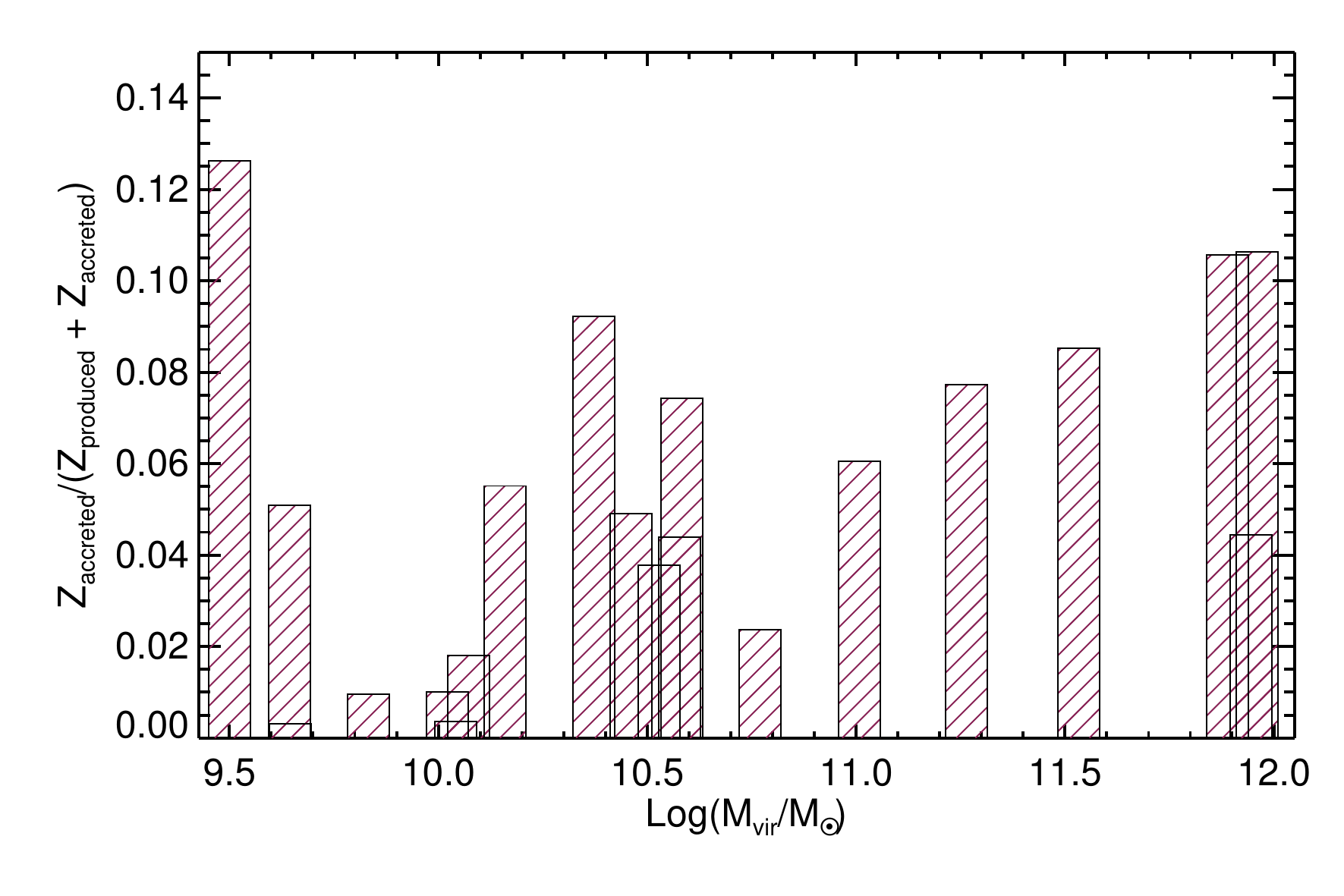}
\end{center}
\caption[Accreted vs formed in situ metals]
{ 
The fraction of metals within $R_{vir}$ at $z=0$ that were accreted externally as gas. 
The denominator includes the sum of all metals produced and all metals accreted as gas from external sources, including through galaxy mergers.
In general, a larger fraction of the metals available to higher mass galaxies are accreted externally but the fraction of metals externally accreted is low across the entire mass range.
}
\label{fig:insituz}
\end{figure}

In the remainder of this section, we further quantify how the amount of material accreted, outflowing, and reaccreted scales with halo mass.
To begin with, we quantify the role of external accretion in contributing metals.
The fraction of metals that were originally accreted from external sources as gas is shown as a function of halo mass in figure~\ref{fig:insituz}.
In this analysis, the mass of metals accreted as gas is compared to the total mass of metals accreted as gas and produced within the main progenitor by $z = 0$.
As also seen in figure~\ref{fig:outflow_history}, the fraction of metals externally accreted is uniformly small.
However, there appears to be a mass trend, with the most massive galaxies generally accreting a larger fraction of their metals from external sources, probably primarily through mergers.

\begin{figure}
\begin{center}
\includegraphics[width=0.5\textwidth]{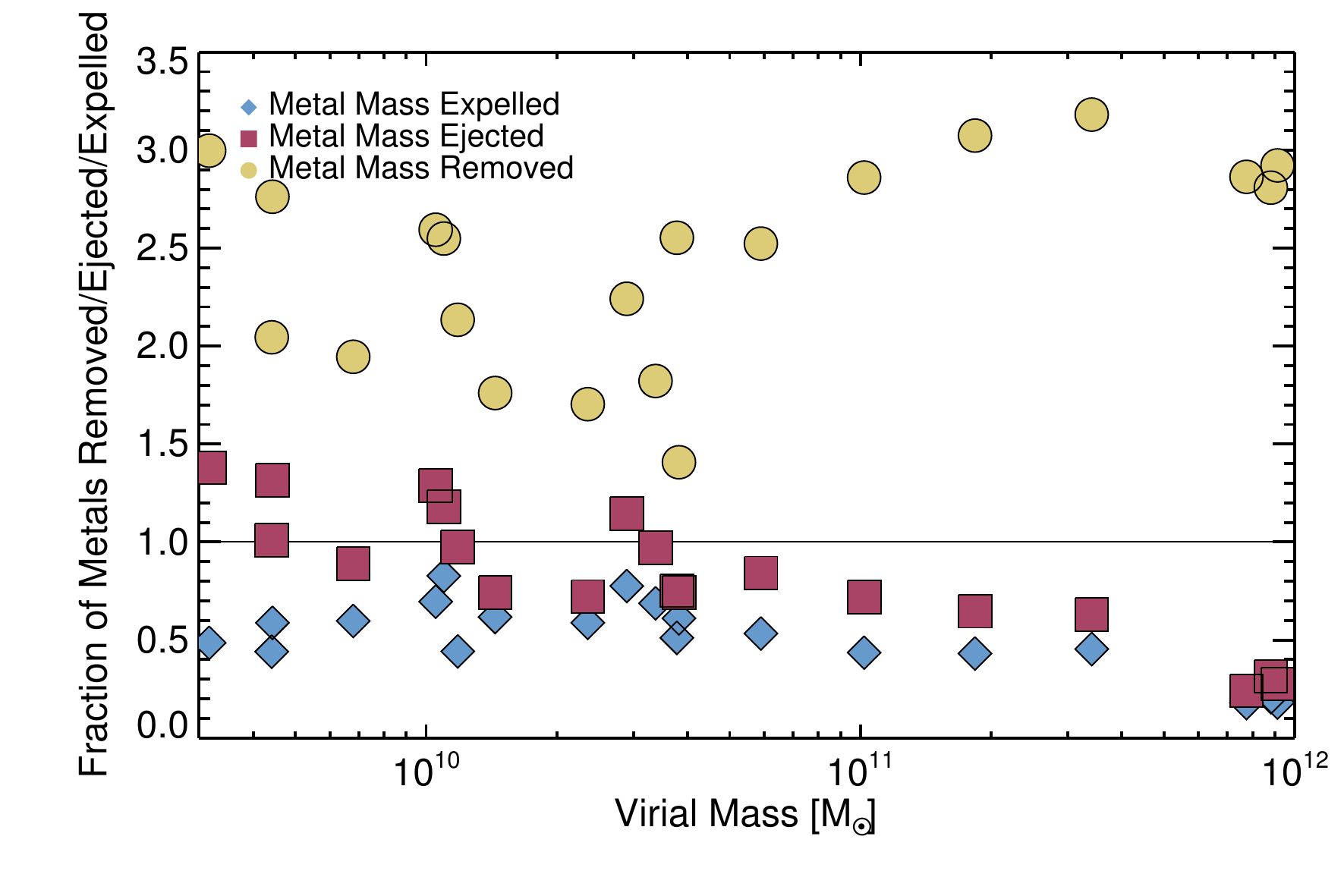}
\end{center}
\caption[]
{ 
Fraction of metals produced or accreted as gas that were removed from the disk (yellow circles), were ejected from the disk (i.e., became dynamically unbound from the disk, maroon squares), or were expelled beyond the virial radius (blue diamonds).
Because metals can be lost again after being reaccreted, this fraction can be greater than one.
}
\label{fig:fraclost}
\end{figure}

Figure~\ref{fig:fraclost} shows the mass of metals in outflows divided by the total mass of metals available (metals either produced by stars or externally accreted as gas).
Because the same metals can exit the disk multiple times, this number is frequently greater than one.
In particular, the large amount of metals removed from the disk compared to the amount available highlight the prevalence of both gas removal and gas reaccretion.
The likelihood of metals being removed from the disk is largely independent of halo mass, because the possibility of gas being heated is basically independent of the galaxy dynamics.
The likelihood of metals being ejected (i.e., part of an outflow), though, is sensitive to mass, because the particles must exceed an energy threshold.
As a result, in more massive galaxies a smaller fraction of either the metals available or the metals removed from the disk are actually considered ejected.
Even in the most massive halos, though, $\sim$ 20\% of the available metals are ejected from the main progenitor.
When considering the fraction of metals expelled beyond the virial radius for galaxies with virial masses greater than $\sim 2.5\times10^{10} \Msun$, we see a similar, although slightly reduced trend as for the mass of metals removed from the disk .
Specifically, the fraction expelled is $\propto$ log(M$_{vir}/\Msun$)$^{-0.31}$ while the fraction ejected is $\propto$ log(M$_{vir}\Msun$)$^{-0.44}$.
As seen in figure~\ref{fig:outflow_history}, in these more massive galaxies, a majority of the gas particles that exit the disk also exit the virial radius, explaining the similarity in the trends.
For galaxies with $M_{vir} \lesssim 3\times 10^{10} \Msun$, however, the fraction of metals that are expelled beyond the virial radius is roughly constant, even as the fraction that exits the disk increases with decreasing halo mass.
As a result, in the lowest mass galaxies a minority of the metals ejected from the disk are able to escape the virial radius.
This effect is especially strong for the three lowest mass galaxies, which have been previously shown to retain metals relatively close to their centers (figures~\ref{fig:metaldist} and ~\ref{fig:cumhistr}).

\begin{figure}
\begin{center}
\includegraphics[width=0.5\textwidth]{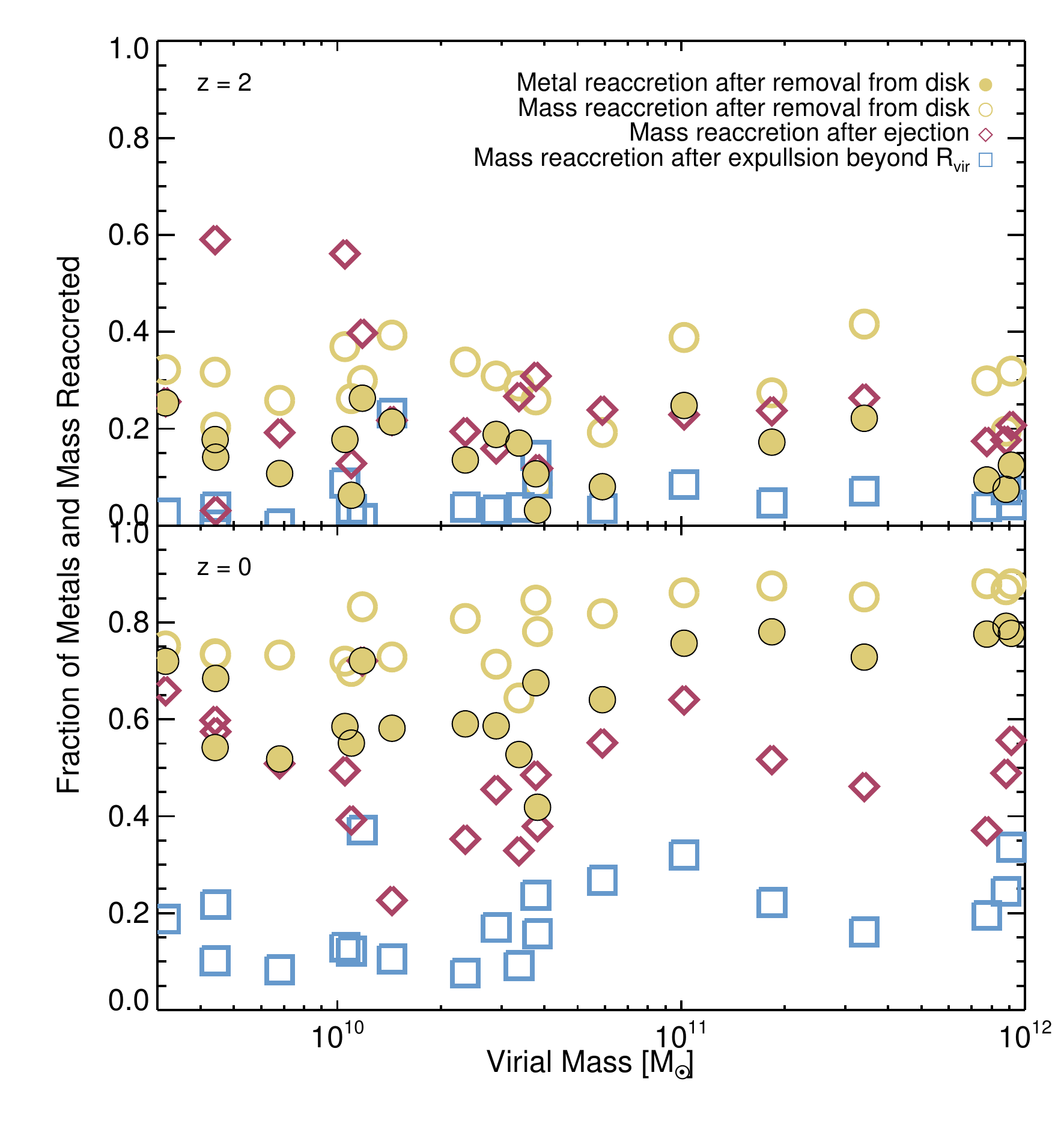}
\end{center}
\caption[Reaccretion fraction]
{ 
Fraction of metals and mass that are reaccreted by z = 2 (top panel) and z = 0 (bottom panel).
Open symbols show the mass fraction of particles reaccreted after either being removed from the disk (yellow circles), being ejected from the disk (maroon squares), or expelled beyond the virial radius (blue diamonds).
Filled yellow circles show the fraction of metal mass removed from the disk that is later reaccreted.
}
\label{fig:fracreaccr}
\end{figure}

Figure~\ref{fig:fracreaccr} quantifies the fraction of gas reaccreted.
As throughout this section, we divide between gas removed from the disk, ejected from the disk, and expelled beyond the virial radius.
The fractions of the gas that are reaccreted by $z = 2$  and $z = 0$ are shown as a function of halo mass.
All reaccretion rates rose from $z = 2$ to $z = 0$ as the greater time elapsed allowed for more material to cycle back.
As anticipated, reaccretion rates are lower for outflowing material satisfying a more stringent energy cut.
However, rates of reaccretion at $z = 0$ are substantial throughout. 
In a couple of galaxies the fraction of mass reaccreted after having left the virial radius even exceeds 30\%.
At $z = 0$, we see slight positive mass trends in the fraction of mass reaccreted after removal from the disk.
The fractions of mass reaccreted after ejection or expulsion do not show consistent trends with mass, likely because the mass of the halo is already incorporated into the determination of whether a particle is ejected or expelled.

We also show the metal fraction reaccreted after having been removed from the disk.
As in figure~\ref{fig:outflow_history}, we determine the mass of metals reaccreted by subtracting the mass of metals externally accreted from the total mass of metals accreted onto the disk.\footnote{We do not show the metal fraction returned by previously ejected or expelled particles because metal diffusion allows metals to be transferred from ejected particles to other particles in the halo.}
The fraction of metals returned, while high, is noticeably lower than the mass fraction returned.
This difference is likely the result of the correspondence between the amount of metals and the amount of energy transferred to gas from SNe.
Those particles least likely to return to the disk are those that received the most SN energy and, presumably, relatively large amounts of metals.

\subsection{Outflow metallicities}\label{sec:outflowmetals}

\begin{figure*}
\begin{center}$
\begin{array}{cc}
\includegraphics[width=1.0\textwidth]{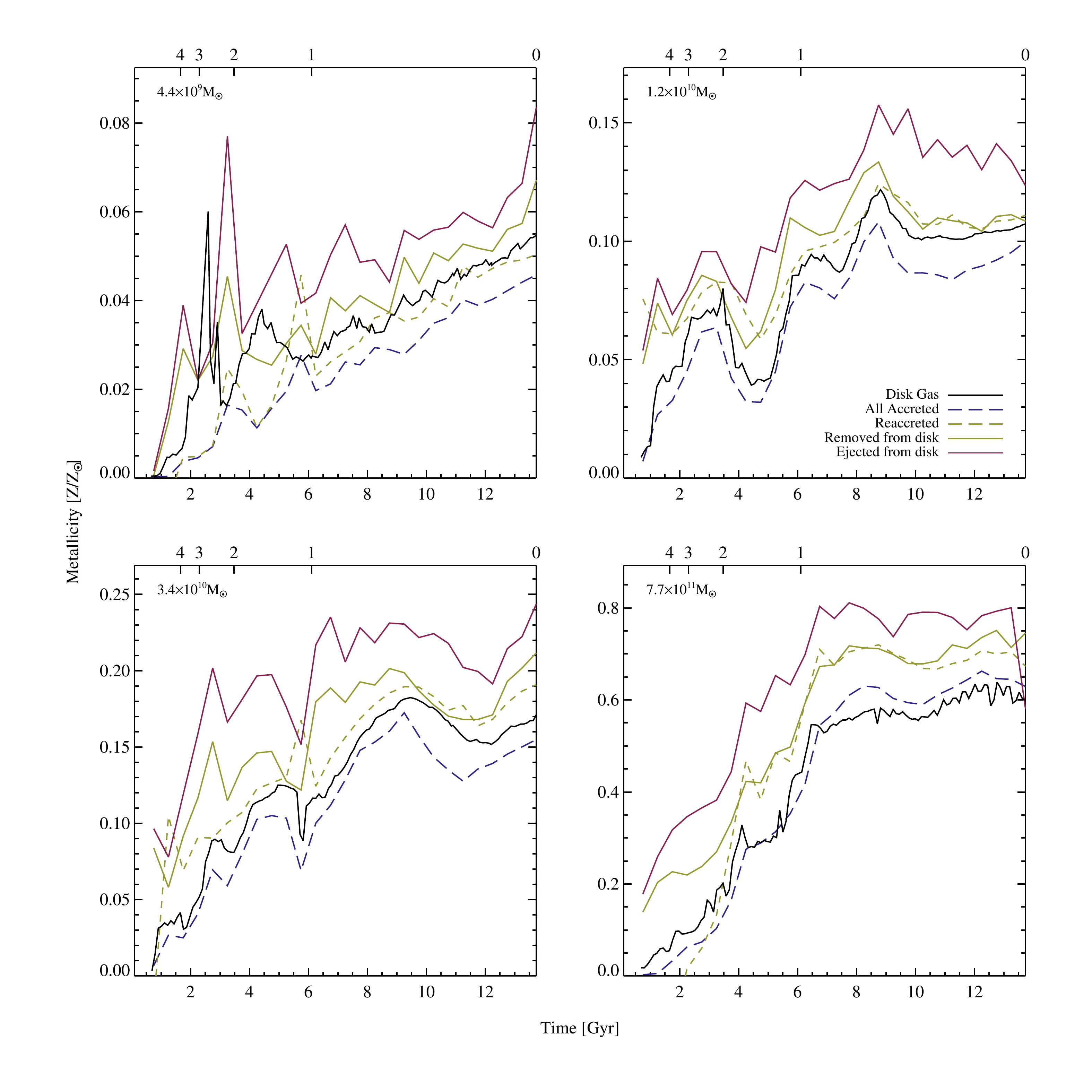}
\end{array}$
\end{center}
\caption[Metallicity of ejected and accreted material]
{ 
History of average metallicities of outflowing and accreting material for four example galaxies spanning a range of masses.
The virial masses of each galaxy are shown in the top left corners of the panels.
The solid black line shows the average metallicity of the ISM.
Solid colored lines show the average metallicity of gas that is removed from the disk (gold) and ejected such that it dynamically escapes the disk (maroon).
The long dashed purple line follows the average metallicity of all accreted and reaccreted material.
The short dashed gold line show the average metallicity of material reaccreted onto the disk.
}
\label{fig:zejecta_gals_history}
\end{figure*}

In analytic models of halo enrichment, outflows are frequently assumed to share the same metallicity as the ISM \citep[e.g.][]{Dave2012, Lilly2013}.
However, the correspondence between supernovae enrichment and the generation of outflows implies that outflows may be metal-enriched compared to the rest of the ISM.
Recent observations of outflows by \citet{Chisholm2016} appear to confirm the relative enrichment of outflows.
If true, the degree of enrichment would be an important parameter in modeling the metallicities of galaxies.
Here, we examine the history of outflow and inflow metallicity in comparison to the ISM.

Figure~\ref{fig:zejecta_gals_history} shows average metallicities of outflowing and accreting material over time for four representative galaxies spanning our range of masses.
Metallicities of both gas removed from the disk and ejected are shown.
Ejected material is more metal rich than material simply removed from the disk because the ``ejection" criteria selects for gas particles that receive sufficient supernovae feedback to dynamically escape the disk and are, consequentially, more likely to also receive large amounts metals.
Both types of outflowing gas, though, are enhanced compared to the ISM, either because of enrichment by the supernovae driving the outflow or because the particles originate in already metal enhanced areas of on-going star formation.

We also show the metallicities of accreted material.
The metallicity of all accreted gas incorporates both the metals accreted from external galaxies and those reaccreted to the disk.
We calculated the metallicity of the subset of that material that is reaccreted by excluding the gas mass that was being accreted to the disk for the first time and those metals that had been accreted onto the halo.
As expected, the reaccreted material is an especially metal-enriched subset of the total accretion.
Both accretion and reaccretion tend to track the metallicities of the disk material, since that is the primary source of metals in the halo.
Occasionally, the metallicity of the reaccreting material appears to be offset in time from the metallicity of the ejected and removed material, e.g. the top left panel between 2 -- 6 Gyrs.
This temporal offset is a clear signature of fountaining.
However, the short reaccretion timescales  of $\sim 1$ Gyr \citep{Christensen2016} and noise in the data make them difficult to identify.

\begin{figure}
\begin{center}
\includegraphics[width=0.5\textwidth]{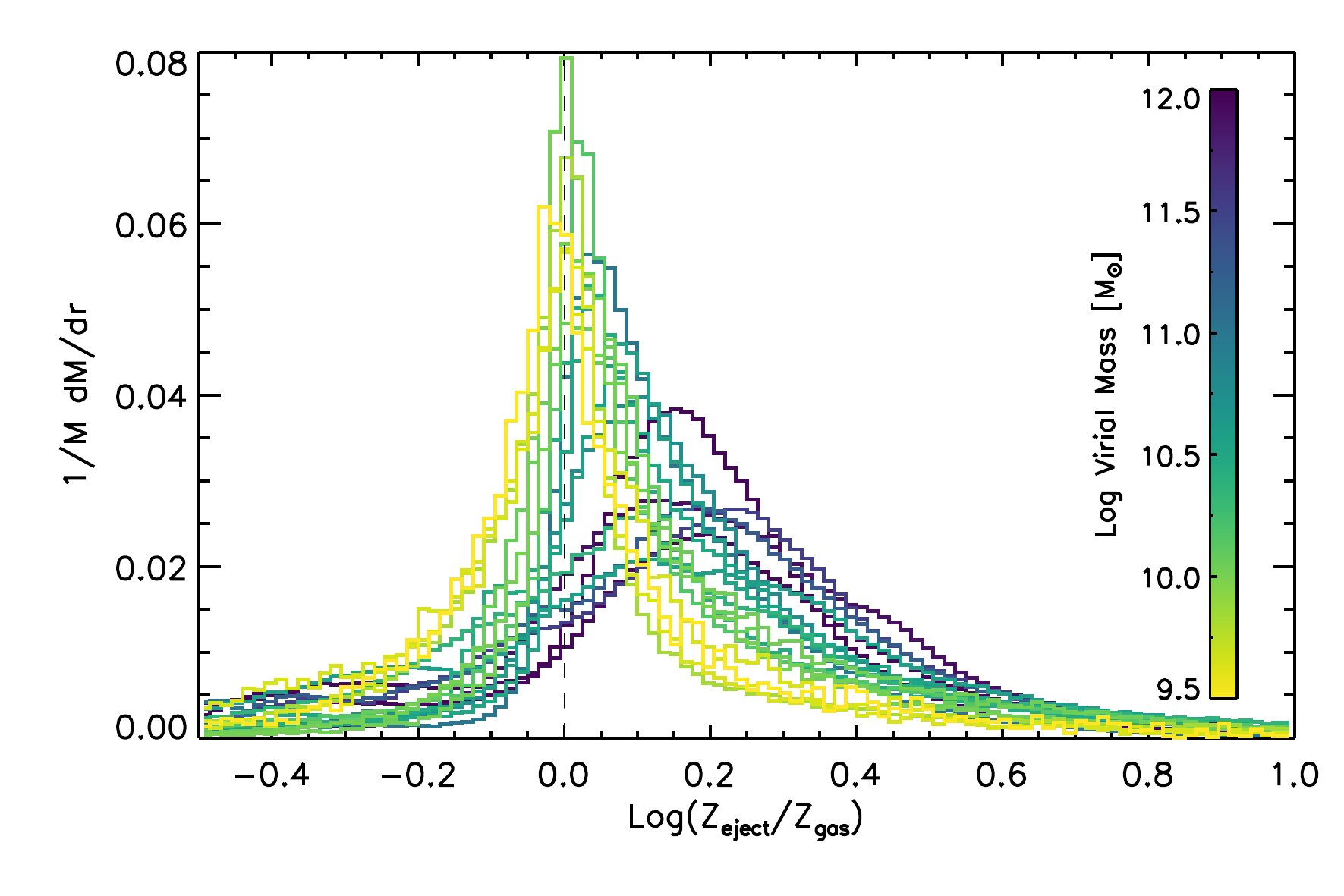}
\end{center}
\caption[Metallicity of ejected material]
{ 
Normalized histogram of the metallicities of ejected gas particles divided by the mean metallicity of the ISM at the time of ejection.
Each curve represents all gas ejected over the history of a single galaxy, with different colors corresponding to different redshift zero virial masses.
Most curves peak close to one with a long tail towards higher metal enrichment.
The curves for more massive galaxies tend to peak at higher levels of metal enrichment, perhaps indicating reduced amounts of entrained material in outflows from those galaxies.
}
\label{fig:zejecta_gals}
\end{figure}

In order to further study mass trends in the relative metal enhancement of outflowing material, we examine the distribution of ejected particle metallicities for all galaxies in figure~\ref{fig:zejecta_gals}.
This figure shows histograms of the relative metal enrichment of the ejected material.
To determine this enrichment, the metallicity of all ejected gas particles at the snapshot after them leaving the disk was divided by the mean metallicity of the ISM at that snapshot.
The histograms tend to peak close to one, indicating that ejected gas is most likely to share a similar metallicity to the ISM.
However, the histograms also show long tails towards higher levels of metal enrichment, which raises the overall average metal enrichment of outflows.
Highly enriched ejected gas particles are most likely the result the simultaneous transfer of large amounts of metals and energy from nearby SNe, while gas particles with metallicities closer to that of the ambient ISM are more likely to have been ejected through entrainment.
The mass trend seen in these histograms may also be explained by differing amounts of entrainment.
Histograms for low mass galaxies peak closer to one than high mass galaxies, probably because larger amounts of ambient ISM are carried out from these galaxies during outflow events \citep{Christensen2016}, although a more homogeneous distribution of metals in the ISM of dwarf galaxies could have a similar effect.

\begin{figure}
\begin{center}
\includegraphics[width=0.5\textwidth]{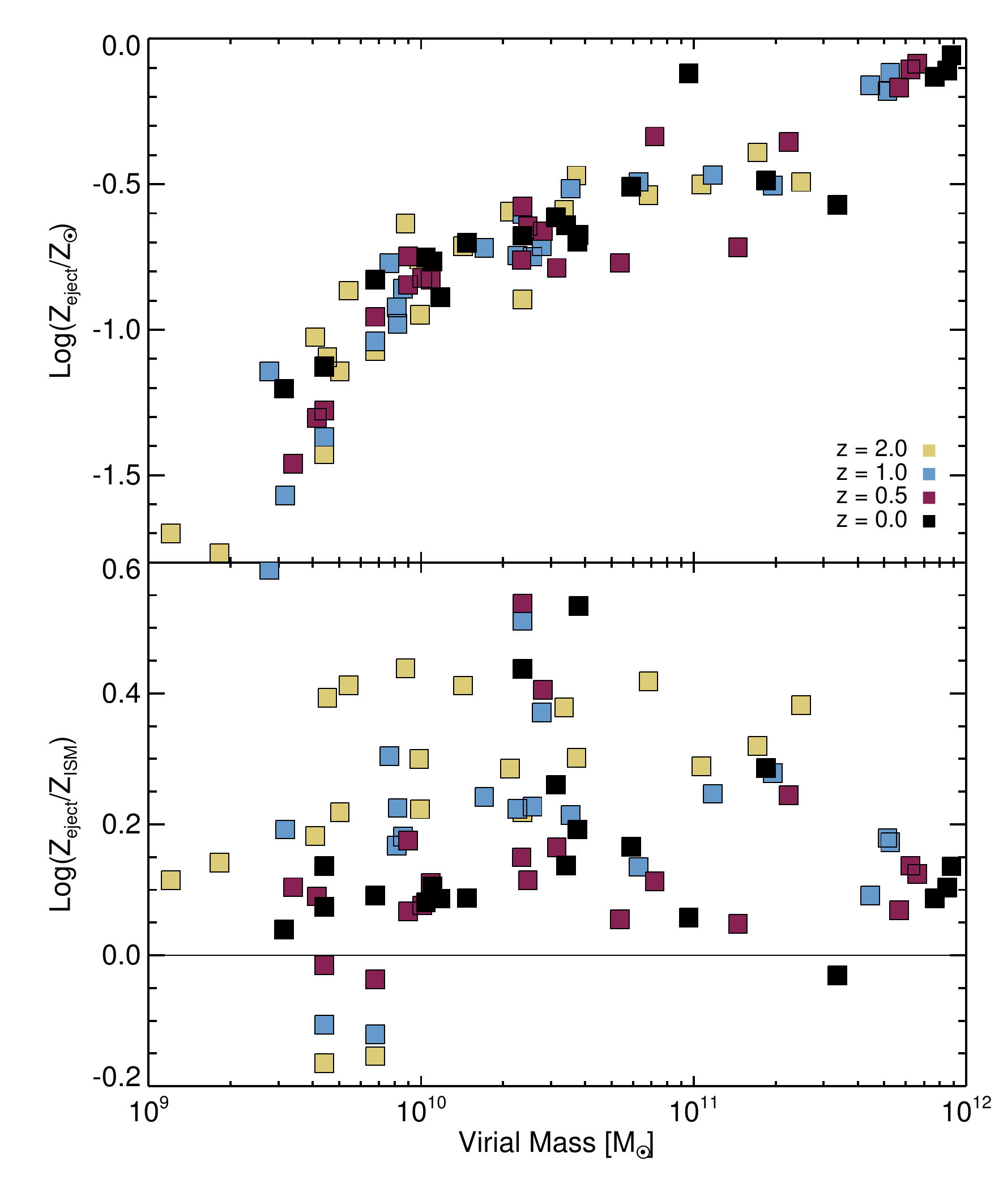}
\end{center}
\caption[Metallicity of ejected material]
{ 
Metallicity of ejected material.
The top panel shows the mean metallicity of the gas ejected at different redshifts as a function of virial mass.
There is a trend towards increasing metallicity with increasing virial mass, as is expected from the MZR.
There is no observed redshift evolution in the relationship between outflow metallicity and galaxy mass.
The bottom panel shows the mean metallicity of the ejected material normalized by the mean metallicity of the disk gas at that redshift.
Compared to the ambient gas, outflows generally involve much more highly enriched material.
The relative enrichment of the ejected material increases with $z$.
}
\label{fig:zejecta}
\end{figure}

We further quantify the mean metallicities of outflows at different redshifts (figure~\ref{fig:zejecta}, top panel).
Unsurprisingly, the ejecta metallicity, like that of the disk gas, increases with halo mass.
There is little to no redshift evolution in the relationship between ejecta metallicity and virial mass despite the evolving mass-metallicity relation. 

Dividing the mean metallicity of the ejected gas by that of the ISM at the time when it was ejected provides a measurement of the relative enrichment of the ejecta at different redshifts (figure ~\ref{fig:zejecta}, bottom panel).
We see a large range of enrichment levels and a few cases where the outflow metallicity is actually lower than the ambient ISM.
We also do not observe a clear metal enrichment trend with mass, although there is some evidence that intermediate mass galaxies have the highest level of metal enrichment.
Conversely, the very lowest mass galaxies had some of the smallest amounts of relative metal enrichment, possibly arising from higher rates of entrainment in these galaxies.
Therefore, the low metallicities of dwarf galaxies are not because they preferentially ejected metals compared to more massive galaxies but are rather because they are more efficient at ejecting material in general, as will be explored in the next section.
Metal enrichment levels did tend to be higher at $z = 2$ and, to a lesser extent, at $z = 1$.
At these redshifts, the ISM metallicity would have been lower, leading to a greater difference between it and the recently-enriched gas near supernovae.

By $z = 0$, the average relative enrichment was only a factor of 1.5.
These redshift zero metal enrichment results for the most massive galaxies are consistent with observations of NGC 6090 (M$_* = 10^{10.7} \Msun$), which showed a factor of 1.3 = $10^{0.11}$ times greater metallicity in outflows than the ISM  \citep{Chisholm2016}. 
Similar, although slightly smaller, amounts of metal enrichment were also found by \citet{Muratov2017} in their simulations. 
Specifically, for redshifts $4 < z < 0$, they found that winds were generally more metal enriched by a factor of $\sim$1-1.5.
As in this work, they also found a trend toward greater metal enrichment at higher redshift. 
Likewise, they did not observe a general mass dependency but did observe that outflows with no metal enrichment (or even metal depletion) compared to the interstellar media came from the smallest galaxies.

\subsection{Metal Mass Loading}\label{sec:massloading}

\begin{figure}
\begin{center}
\includegraphics[width=0.5\textwidth]{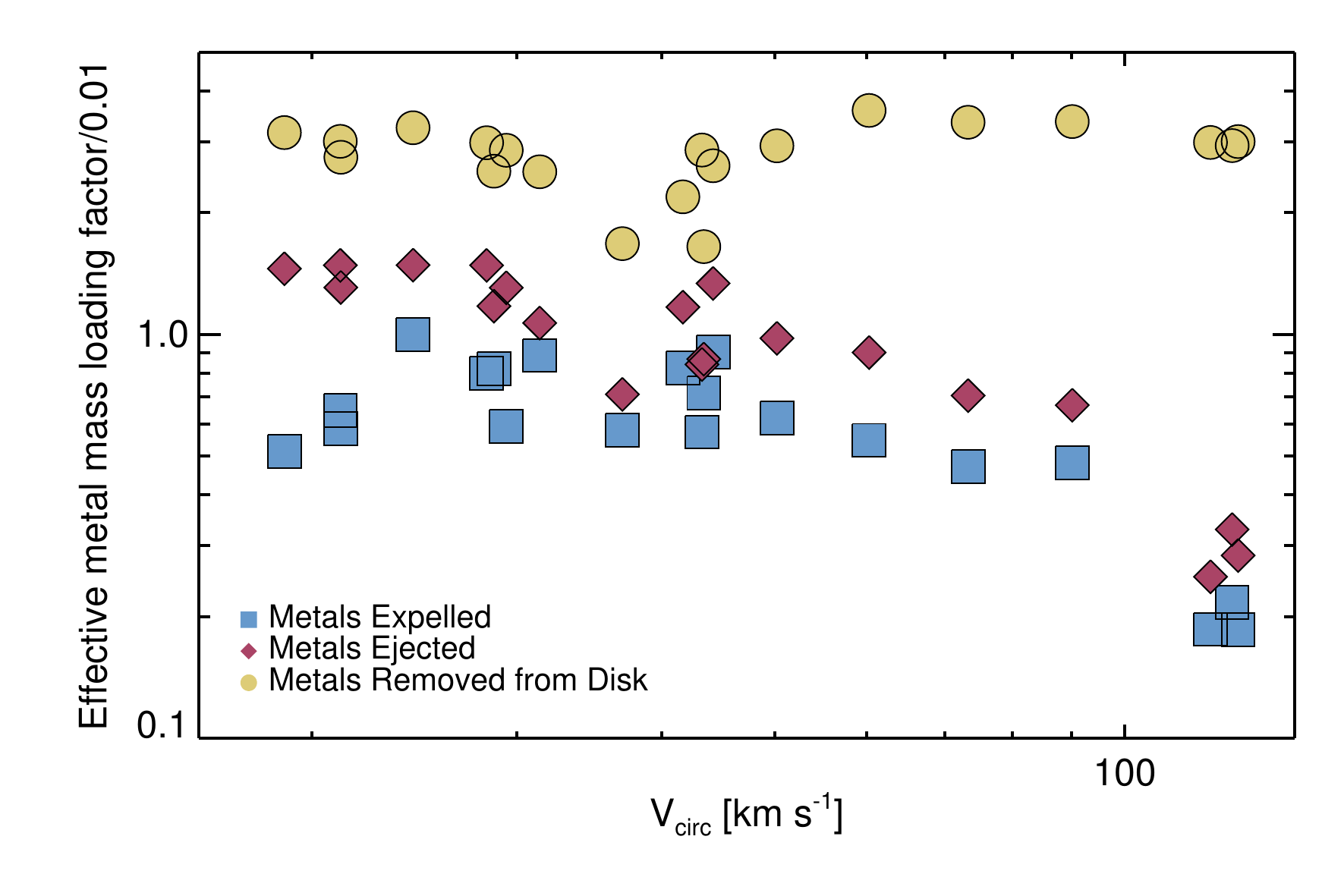}
\end{center}
\caption[Metal mass loading, effective]
{ 
The total mass of metals lost over the history of the galaxy normalized by the total stellar mass formed, also known as the effective metal mass loading, shown as a function of galaxy circular velocity.
The effective metal mass loading shown here is scaled by the stellar yield for the \citet{Kroupa93} IMF: 0.01. 
Different symbols represent different selection criteria for outflowing material: gold circles include all material that leaves the disk, maroon squares show only ejected material that exceeds the escape velocity of the disk, and blue diamonds show only that material that eventually is expelled beyond the virial radii.
Tightening the criteria for outflow identification to require gas to either dynamically escape the disk or leave the halo both reduces the effective metal mass loading and introduces a mass dependency.
While similar amounts of metals are removed from the ISM because of stellar feedback, the deeper potential wells of higher mass galaxies result in a smaller mass of metals in outflows per stellar mass formed.
}
\label{fig:zml_eff}
\end{figure}

The efficiency of galaxies at expelling their metals can be quantified as the ``metal mass loading" factor, i.e. the rate at which metals are ejected divided by the rate at which star formation occurs.
An effective metal mass loading representing the metal loss integrated over the history of the galaxy can be found by dividing the total mass of metals in outflows by the total mass of stars formed.
Figure~\ref{fig:zml_eff} shows the effective metal mass loading as a function of circular velocity for several different ways of identifying outflows: all gas removed from the disk, gas that exceeds the escape velocity for the disk (``ejected"), and gas that is expelled beyond the virial radius.
The metallicities used in these calculations are the metallicities of gas particles at the snapshot immediately prior to their removal.
As would be expected, increasingly stringent criteria for identifying outflowing material results in lower effective metal mass loading rates.
Requiring that the outflowing material satisfy an energy criteria, either exceeding the escape velocity or leaving the virial radii, also introduces a mass dependency to the effective metal mass loading.
While similar amounts of metals per stellar mass formed are removed from the disk across the entire mass range of galaxies, the deeper potential wells of more massive galaxies result in smaller fractions of the metals able to dynamically escape the disk.
The result is substantially lower efficiencies of metal loss through outflows in more massive galaxies.

\begin{figure}
\begin{center}
\includegraphics[width=0.5\textwidth]{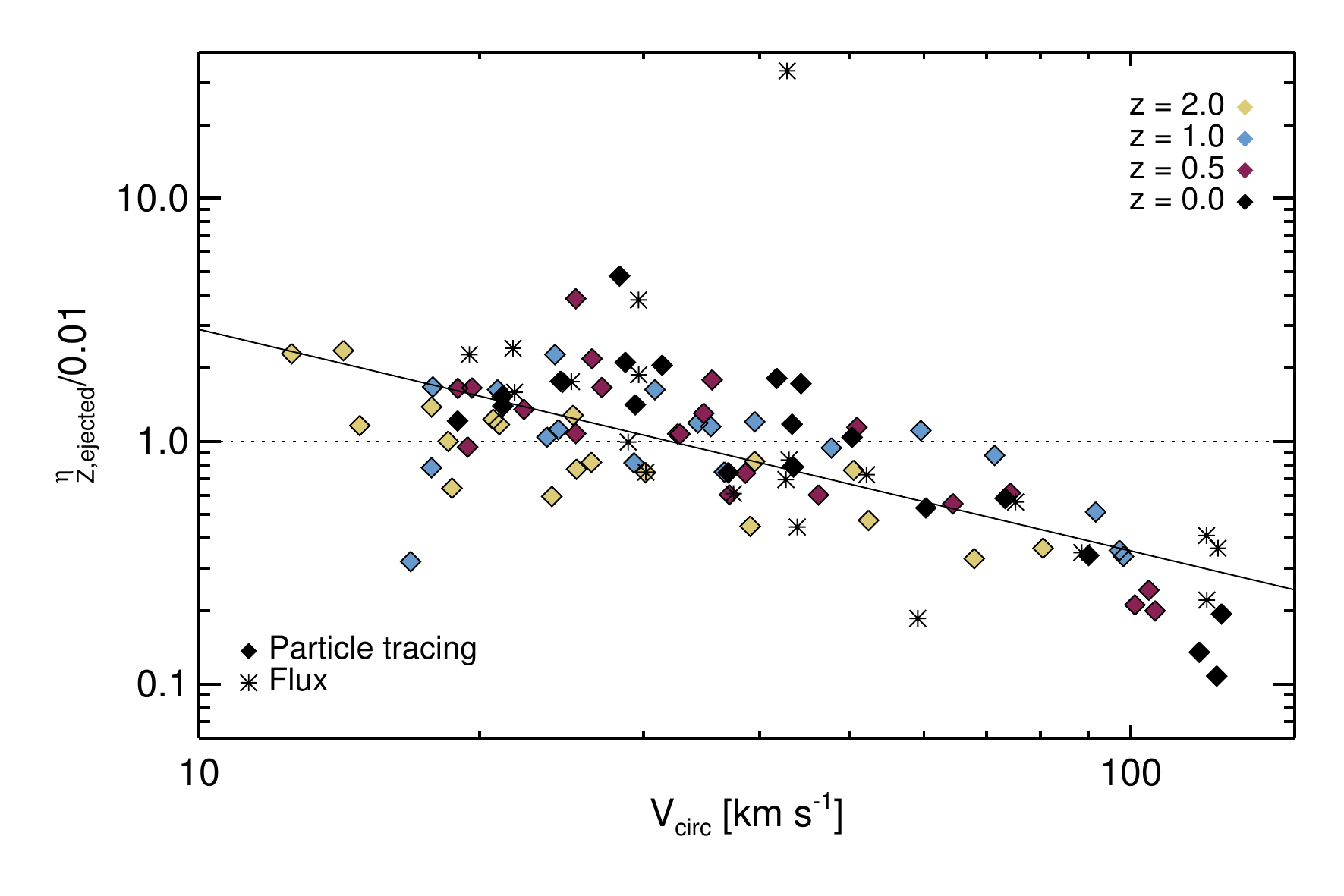}
\end{center}
\caption[Metal mass loading, instantaneous]
{ 
The mass of metals ejected in particle tracking-selected outflows divided by the mass of stars formed in 1 Gyr time bins (i.e., the instantaneous metal mass loading) as a function of galaxy circular velocity (filled symbols).
Measurements were made at $z$ = 0, 0.5, 1 and 2.
A logarithmic fit to the ejection data at all redshifts is shown by the solid line.
The asterisks show the instantaneous metal mass loading at $z = 0$ as calculated by measuring flux through a spherical shell of radius $0.25 R_{vir}$, for comparison.
The instantaneous metal mass loading factors shown here are scaled by a typical stellar yield for a \citet{Kroupa93} IMF: 0.01.
}
\label{fig:zml_instant}
\end{figure}

Figure~\ref{fig:zml_instant} shows a more instantaneous metal mass loading factor, $\eta_{metals}$, as determined using particle tracing-selected outflows (filled symbols) and through a calculation of the metal flux (asterisks).
In the case of the particle tracing selected outflows, metal loss and star formation rates are calculated for 1 Gyr time bins centered on each of four redshifts (z = 0, 0.5, 1.0, and 2.0).
We only consider the metals carried by gas exceeding the disk escape velocity (``ejected"), as this criterion best selects for gas generally considered part of outflows.
A power-law fit to the data for all redshifts results in $\eta_{metals} \propto v_{circ}^{-0.91}$.
A similar fit to the total mass loading function in \citet{Christensen2016} had a $v_{circ}^{-2.2}$ dependency.
The shallower dependency of the metal mass loading is the result of the MZR.
The relation can be explained by scaling the mass loading relation by the metallicity of the outflow.
As seen in figure~\ref{fig:zejecta}, outflow metallicity increases with virial mass following a scaling of approximately $Z_{ejecta} \propto M_{vir}^{0.46} \propto v_{circ}^{1.4}$.
This scaling is very similar to that observed for the low-end of the MZR, as would be anticipated from the lack of mass scaling in the relative metal enhancement of the ejecta.
Therefore, $\eta_{metals} = Z_{ejecta} \eta_{total} \propto v_{circ}^{1.4}v_{circ}^{-2.2} \sim v_{circ}^{-0.91}$.
As a result, while dwarf galaxies have low metallicities and they do not preferentially eject metals compared to higher mass galaxies, their tendency to eject more mass overall results in high metal mass loading factors.

Metal mass loading factors were previously measured in \citet{Muratov2017} by calculating the metal flux through a $0.25 R_{vir}$ sphere surrounding the center of the galaxy.
In order to draw a direct comparison to their work, we also show the metal mass loading factor as determined from the flux, $\eta_{flux}$, following the method in \citet{Muratov2017}.
Specifically, we identify all particles with outward radial velocities within a spherical shell of inner radius 0.2 $R_{vir}$ and outer radius 0.3 $R_{vir}$ as part of an outflow.
Using these particles, $\eta_{Z, flux}$ is defined to be
\begin{equation}
\eta_{Z, flux} = \frac{1}{\dot M_{SFR}} \frac{\partial M}{\partial t} = \frac{1}{\dot M_{SFR}} \Sigma v_{rad} Z_{SPH} m_{SPH}/dL
\end{equation}
where $M_{SFR}$ is the star formation rate averaged over 100 Myr, $v_{rad}$ are the radial velocities of the outflowing gas particles, $Z_{SPH}$ are the metallicity of the particles, $m_{SPH}$ are their masses, and dL is the width of the spherical shell (0.1 $R_{vir}$).
Similarly to \citet{Muratov2017}, we calculate $\eta_{Z, flux}$ at five different snapshots in the last 3 Gyrs to reduce the noise.
We find that calculating the metal mass loading factor using flux through a sphere, rather than particle tracing, introduces a greater amount of scatter.
However, the trend and scaling is similar for both methods, indicating that our metal mass loading factors are largely insensitive to the method used to measure them.

\citet{Muratov2017} found a metal mass loading factor approximately equal to the SNe II yield (0.02) for their simulations across all masses and redshifts, which they interpret as implying that all metals produced through Type II supernovae are immediately ejected, at least temporarily.
In contrast, we find a definite mass dependency.
We calculate metal mass loading factors for dwarf galaxies slightly higher than simulation yields and metal mass loading for Milky Way-mass galaxies almost a factor of five lower than the yields.
The slightly higher metal mass loading factors than yields in our dwarf galaxies likely indicates that entrainment is removing (at least temporarily) additional metals beyond those produced by the SNe, while the lower metal mass loading factors for more massive galaxies demonstrates that supernova are incapable of removing most of the produced metals from the more massive disks.
Both of these phenomena are likely absent in the \citet{Muratov2017} FIRE simulations.

\section{Discussion}\label{sec:discuss}
\subsection{Generation of the mass-metallicity relation}
The MZR can be explained through a combination of inflow, outflow, recycling rates, in addition to star formation efficiencies.
Here we discuss the role of these processes in establishing the MZR in the simulated galaxies presented here.
As a framework, we modify the analytic model outlined in Appendix C of \citet{Peeples2011} to include reaccretion \citep[see][for a similar modification]{Zahid2014}.

One can write the instantaneous change in the metal mass, $\dot M_Z$, as 
\begin{equation}
\begin{split}
\dot M_Z = Z_{IGM} \dot M_{acc} - Z_{ISM} \dot M_{SFR} + Z_{ej}\dot M_{recy}  \\
-  Z_W \dot M_W + Z_{reaccr}\dot M_{reaccr}\\
\label{eq:mdot_1}
\end{split}
\end{equation}
where $Z_{IGM}$ is the metallicity of the IGM, $\dot M_{acc}$ is the rate of external mass accretion, $Z_{ISM}$ is the metallicity of the ISM, ${\dot M_{SFR}}$ is the star formation rate, $Z_{ej}$ is the metallicity of gas being returned to the ISM by stars, $\dot M_{recy}$ is the rate at which mass is returned from ISM by stars,  $Z_W$ is the metallicity of the galactic wind, ${\dot M_W}$ is the rate of mass outflow, $Z_{reaccr}$ is the metallicity of the reaccreted material, and ${\dot M_{reaccr}}$ is the rate of mass reaccretion.
We can define a nucleosynthetic yield: 
\begin{equation}
y =  Z_{ej} \frac{\dot M_{recy}}{\dot M_{SFR}} = Z_{ej}f_{recy}. 
\label{eq:yield}
\end{equation}
For the IMF and theoretical yield models used in this paper, $y \approx 0.01$.
By also defining a metal accretion efficiency, $\zeta_a =  \frac{Z_{IGM}}{Z_{ISM}}\times\frac{\dot M_{acc}}{\dot M_{SFR}}$, 
equation~\ref{eq:mdot_1} can be simplified to 
\begin{equation}
\begin{aligned}
\dot M_Z & = \dot M_{SFR} \\
& \left( y + Z_{ISM}\left( \zeta_a - 1 - \frac{Z_W \dot M_W - Z_{reaccr} \dot M_{reaccr}}{Z_{ISM} \dot M_{SFR}}\right) \right).
\end{aligned}
\label{eq:mdot}
\end{equation}
Similarly to $\zeta_a$, we define a net metal expulsion efficiency.
\begin{equation}
\zeta_{net} = \frac{Z_W \dot M_W - Z_{reaccr} \dot M_{reaccr}}{Z_{ISM} \dot M_{SFR}}
\end{equation} 
Provided reaccretion happens on reasonably short time periods, $\dot M_{reaccr} = f_{reaccr} \dot M_{W}$, and 
\begin{equation}
\zeta_{net} = \frac{\dot M_W (Z_W  - Z_{reaccr} f_{reaccr})}{Z_{ISM} \dot M_{SFR}}
\end{equation} 
where $f_{reaccr}$ is the fraction of material leaving the disk that is later reaccreted. 
If there were to be no reaccretion of wind material, $f_{reaccr}$ would be equal to 0 and $\zeta_{net}$ would reduce to $\zeta_{net} = \frac{Z_W}{Z_{ISM}}\times\frac{\dot M_W}{\dot M_{SFR}}$.
By substituting in $\zeta_{net}$ to equation~\eqref{eq:mdot}, it reduces to
\begin{equation}
\dot M_Z = \dot M_{SFR} \left( y + Z_{ISM}\left( \zeta_a - 1 - \zeta_{net}\right) \right)
\end{equation}

The scaling of $\zeta_{net}$ with halo mass uniquely determines the MZR if a stellar mass-halo mass relation is adopted and $y$ and $\zeta_a$ are assumed to be independent of halo mass.
To simplify further, one could assume that $\zeta_a \approx 0$, an assumption supported by the minimal metallicity of externally accreted material measured in our simulations (figure~\ref{fig:outflow_history}).
Similarly, $y$ is independent of time for a given IMF. 

By combining equations governing the instantaneous change in total mass and metal mass and assuming a power law relationship between gas fraction and stellar mass such that $F_g = \frac{M_g}{M_*} = K_f M_*^{-\gamma}$, one can write
\begin{equation}
\frac{\mathrm{d}Z_g}{\mathrm{d}Z_*} = \frac{y + Z_g(\zeta_a - 1-  \zeta_{net} - F_g(1 - \gamma))}{M_g(1 - f_{recy})}
\end{equation}
following the derivation in Appendix C of \citet{Peeples2011}.
For a given relationship between $\zeta_{net}$ and $M_*$, this equation can be integrated with respect to $M_*$ to produce a MZR.

\begin{figure}
\begin{center}
\includegraphics[width=0.5\textwidth]{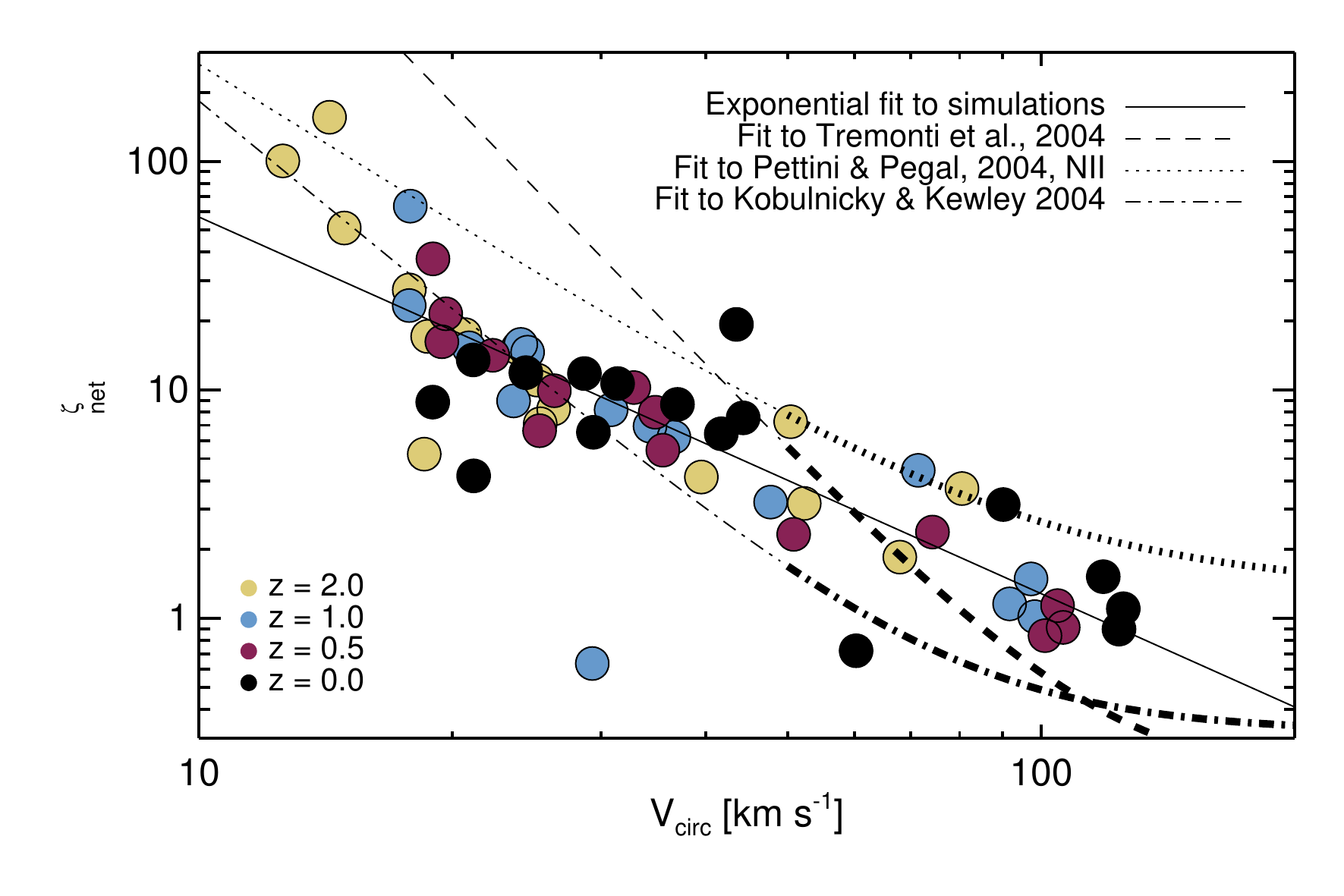}
\end{center}
\caption[Zeta W]
{ 
Net metal expulsion efficiency, $\zeta_{net}$, calculated by dividing the net mass loss from the disk by the total stellar mass formed and scaling by the ISM metallicity and star formation rate, versus the circular velocity of the galaxy.
Measurements are shown for 1 Gyr bins at four different redshifts.
The solid black line shows an power law fit to the data.
Thick patterned black lines show the predicted relationship between $\zeta_W$ and $v_{circ}$ from \citep{Peeples2011}, assuming observed MZRs based on three different indicators, \citet{tremonti04} (not adjusted, dashed ), \citet{Pettini2004} (dotted), and \citet{Kobulnicky2004} (dot-dashed).
The thin patterned black lines show the extrapolation of the predicted relationships below the observed range.
}
\label{fig:zetaw}
\end{figure}

In figure~\ref{fig:zetaw}, we show $\zeta_{net}$ versus $v_{circ}$ for our simulations.
To calculate $\zeta_{net}$, we determined the {\em net} mass removed from the disk across 1 Gyr time bins at four different redshifts and scaled the values by the ISM metallicity and star formation rate at those times.
Excluding the three out of 40 instances when the metal influx exceeded the metal outflux, we find that the data is well fit by a power law: $\zeta_{net} = 3.4 v_{circ}^{-1.7}$.
The scaling with $v_{circ}$ is less extreme than the mass loading factor, $\eta_{total} \propto v_{circ}^{-2.2}$, but more so than the metal mass loading factor, $\eta_{metals} \propto v_{circ}^{-0.91}$.
As  $f_{reaccr}$, the fraction of material reaccreted after leaving the disk, is relatively constant  with halo mass (figure~\ref{fig:fracreaccr}), the more extreme mass scaling for $\zeta_{net}$ than $\eta_{metals}$ must arise from the additional factor of $1/Z_{ISM}$ in $\zeta_{net}$.
Since $Z_{ISM}$ is smaller in low-mass galaxies, this factor results in $\zeta_{net}$ having a stronger mass dependency.

For comparison, we also show three best-fit functions for $\zeta_{net}$ to different observed MZRs from \citet{Peeples2011}.
These functions were based on SDSS data of local, star forming galaxies, compiled and adjusted for different metallicity indicators by \citet{Kewley08}.
The indicators, \citet{tremonti04}, \citet{Pettini2004} NII, \citet{Kobulnicky2004}, were chosen to represent the range of possible values.\footnote{Note that in comparing our data to the MZR in \S\ref{sec:mzr}, we showed data using or adjusted to \citet{Pettini2004} NII when possible.}
The fits shown have been extrapolated (shown in a thinner line) beyond the M$_* = 10^{8.5} \Msun$ ($v_{circ} \sim 50$ km s$^{-1}$) limit used in \citet{Kewley08}.

For galaxies with $v_{circ} > 50$ km s$^{-1}$, $\zeta_{net}$ lies within what would be expected for different metallicity indicators, although notably slightly lower than the \citet{Pettini2004} NII calibration we compared to previously in this paper.
For less massive, galaxies, $\zeta_{net}$ is lower than what would be expected by extrapolating fits from more massive galaxies.
However, a slightly flatter MZR at these masses is consistent with data from observed dwarf galaxies by \citet{Lee06}, explaining why the simulations are still able to match the observed MZR at lower masses.
This analysis illustrates the role of outflows in our simulations in setting the MZR.

\subsection{Comparison with other simulations}
The net metal expulsion efficiency, $\zeta_{net}$, encapsulates a large amount of physics that together define the rate of metal loss from the disk for a given halo mass, including the mass loading factors, rates of recycling, and the relative metal enrichment of outflow and reaccretion.
Other simulations with different outflow mechanics may achieve similar fits to the MZR by having similar values of  $\zeta_{net}$.
For example, higher rates of metal outflows may be balanced by higher rates of metal recycling.
This can happen either because of a greater total amount of recycling or because of a smaller difference between the inflow and outflow metallicity. 
Alternatively, higher enrichment of outflowing material relative to the ISM could compensate for lower mass loading factors.
Additionally, different mass scalings of the mass loading factors, amount of reaccretion, and relative enrichment of outflows may conspire to produce similar scalings for $\zeta_{net}$.
We discuss here the numerics of our simulations in comparison to different feedback models and highlight the measurements that could help break the degeneracy.

The delayed-cooling, blastwave model for SNe feedback used in these simulations naturally produces energy-driven wind scalings \citep{Christensen2016}, which is consistent with those models generally assumed for lower halo masses in simulations with hydrodynamically decoupled winds \citep{Ford13b,Vogelsberger2014}.
However, compared to simulations using hydrodynamically decoupled winds, our simulations result in higher amounts of preventative feedback and lower rates of recycling, presumably because of the higher temperature of the halo gas \citep{Christensen2016}.
For instance, \citet{Oppenheimer08} found that about 85\% of wind particles were reaccreted with recycling times scales between $10^{10}$ and $10^{8}$ years, depending on halo mass \citep{Oppenheimer10}, while the blastwave feedback model used here resulted in about half the outflow gas reaccreting at least once with timescales of $10^{9}$ years, independent of halo mass \citep{Christensen2016}. 
Given these differences, we would anticipate hydrodynamically decoupled wind simulations that also reproduce the MZR to have higher values of $\dot M_W$ and $\dot M_{reaccr}$ than our simulations.

Our simulations behave more similarly to those presented in \citet{Muratov2017}, which used a feedback model based on energy and momentum injection according to stellar population synthesis models.
Those simulations had reaccretion times of a few hundred million years for all halo masses \citep{AnglesAlcazar2016}.
While those recycling times are shorter than the $10^{9}$ year timescales measured for our sample of simulations, the similar lack of a mass dependency is striking.
As expected from the shorter recycling times, the fraction of outflowing mass reaccreted was also higher: 50 to 80\% of gas, compared to the 25 to 65\% of ejected material reaccreted in our simulations.
The simulations in \citet{Muratov2017} also have generally higher mass loading factors with a steeper mass dependency.
Their degree of wind metal enrichment is lower and has a shallower mass dependency than in our simulations.
Those two factors combined result in a flat metal mass loading factor compared to the $\eta_{metals} \propto v_{circ}^{-0.91}$ scaling presented here.
We would expect the higher reaccretion rates they find to be balanced by their higher metal mass loading factors, resulting in similar values of $\zeta_{net}$ to us.
However, while some of the difference in reaccretion is almost certainly the result of the different feedback models, these studies do not select outflows the same way, introducing uncertainty in making the comparison.

The range of numerical models able to reproduce the MZR within the observational uncertainties raises the question of what observations could be used to distinguish between them.
The rates and metallicities of inflowing and outflowing material are one set of key quantities; however, selecting the same gas in simulations that observers detect is not trivial.
The amount, temperature, and density of metals in the CGM can provide a complementary constraint to the MZR. 
For example, the distance metals are spread to varies depending on feedback models and should be closely related to the timescales and likelihood of reaccretion.
However, this comparison it limited both by the difficulty in making and interpreting observations of the CGM and the difficulty in creating a simulation that can correctly reproduce the multi-phase CGM.
Furthermore, typical simulations do not resolve the low-density CGM to high levels, so it is possible that substantial amounts of the CGM could be in structures below the resolution limit.

\section{Conclusions}\label{sec:conclude}
In this paper, we analyze the drivers of the metal distribution in a set of twenty simulated galaxies spanning two and a half orders of magnitude in halo mass that match observed characteristics, including the stellar and gaseous MZR.
We follow the accumulation of metals in galaxies by tracking their metal production and by identifying instances of metal accretion and loss.
This analysis enables us to determine the role of galactic winds and reaccretion in determining the mass of metals within different components of a galaxy and its CGM.

\begin{enumerate}
\item Gas outflows are highly effective at removing metals from galactic disks. 
In the lowest mass galaxies, as little as 10\% of the metals produced may remain within the ISM or stars at $z = 0$, while that fraction rises to as much $\sim 50$\% for Milky Way-mass galaxies.
This mass trend is dominated by the stars.
While the ISM retains a similar range of fractions (between 5 and 25\%) of the produced metals across galaxy mass, the fraction locked in stars rises steeply with galaxy mass.
Those metals that do exit the disk of the galaxy are widely dispersed with the majority lying beyond the virial radius. 
Because of their deeper potential wells, more massive galaxies are generally better able to retain their metals within their virial radius and show lower dispersal of metals.
However, this mass trend becomes complicated for dwarf galaxies, as the very lowest mass galaxies are able to retain a moderate fraction of their metals, despite their shallow potentials.
In these galaxies, extremely low star formation rates are likely responsible for the reduced metal loss.

\item The history of metal enrichment of the ISM and stars largely follows the history of metal production by stars with less than 10\% of metals at $z=0$ coming from externally accreted gas or stars.
Large amounts of metals cycle rapidly in and out of the disk with the cumulative history of gas loss sometimes exceeding the mass of metals produced by a factor of three.
The majority of these metals (generally between 50 and 80\%), however, are quickly returned to the disk.
The fraction of these metals that become dynamically unbound from the disk (``ejected") is much higher in low-mass than high-mass galaxies.
These ejected metals are more likely to permanently remain outside of the disk.

\item 
Ejected material tends to be somewhat more metal rich than the ambient ISM, because gas that receives energy from SNe most likely also received metals from the same stellar population.
This effect is largely independent of galaxy mass and is generally stronger at high redshift when the ISM metallicity at a given virial mass would have been lower. 
Gas that is removed from the disk, but not necessarily considered part of an outflow, also shows some amount of metal enrichment, especially at early times, but the effect is reduced, presumably because this material includes gas not as strongly heated by SNe.
Because of dilution by low-metallicity external accretion, the average metallicity of accreted material is lower than outflowing material and, in most cases, the ISM.
However, even when excluding external accretion, the metallicity of material reaccreted after being removed from the disk is lower than the metallicity of the removed material.
This difference in metallicity indicates a tendency for some highly enriched material to remain outside the disk.

\item The metal mass loading factor, $\eta_{metals} = Z_{W} \frac{\dot M_W}{SFR}$, shows a power-law dependency on virial velocity, similar to the standard mass loading factor.
However, we observe a flattening of the power from $a = -2.2$ to $a = -0.91$ when comparing the metal mass loading to the standard mass loading.
This flattening can be entirely explained by the reduced metallicity of the ISM in lower mass galaxies.
Therefore, while low mass galaxies have low metallicity ISM and are no more likely to preferentially eject metals than higher mass galaxies, their exceptionally high mass loading factors still produce high rates of metal loss per stellar mass formed.

\item The MZR can be explained using an alternative depiction of the metal expulsion efficiency, $\zeta_{net}=   \frac{Z_W \dot M_W - Z_{reaccr} \dot M_{reaccr}}{Z_{ISM} \dot M_{SFR}}$, which scales the net metal loss rate from winds by the star formation rate and ISM metallicity.
We find that $\zeta_{net}$ scales as $v_{circ}^{-1.7}$, which is consistent with what would be expected from the observed MZR.
We find the outflowing material to be enriched to the same degree relative to the ISM independent of stellar mass, while reaccretion rates of metals removed from the disk are only slightly higher in more massive galaxies.
As a result, the metal expulsion efficiency, $\zeta_{net}$ shares a similar, though slightly shallower, scaling with $v_{circ}$ as the mass loading factor.
\end{enumerate}

These results illustrate how outflows and gas recycling, in combination with accretion and varying star formation efficiency, can together produce the MZR.
The simulations naturally produce the mass-loading factors of energy-driven winds, with slight metal enhancement and moderate rates of reaccretion across all halo masses.
However, this is not a unique solution to producing the MZR, and further comparisons to the metal content, distribution, and thermodynamic structure of the CGM are necessary to constrain how exactly gas transfer between the ISM and CGM determines the baryonic and metal content of galaxies.

\section{Acknowledgements}\label{sec:acknowledge}
The authors are grateful to the referee for the thoughtful critique that improved this paper.
The authors would like to thank Rachel Somerville for her insightful feedback.
Resources supporting this work were provided by the NASA High-End Computing (HEC) Program through the NASA Advanced Supercomputing (NAS) Division at Ames Research Center as well as the Texas Supercomputing Center. 
TQ acknowledges support from NSF grant  NSF AST-1514868.
RD acknowledges support from the South African Research Chairs Initiative and the South African National Research Foundation.
This research was also supported in part by the NSF under Grant No. NSF PHY11-25915. 

\bibliographystyle{apj} \bibliography{./outflowsZ}

\end{document}